\documentclass[a4paper,11pt]{article}
\pdfoutput=1 

\usepackage{jheppub} 

\usepackage[T1]{fontenc} 

\usepackage{url}
\usepackage{setspace}
\usepackage{amsmath,amssymb,amscd, amsthm}
\usepackage{bbold}
\usepackage{amsfonts}
\usepackage{graphicx}
\usepackage{subcaption}
\usepackage{comment}

\usepackage{bm}
\usepackage{tikz}

\makeatletter
\def\tikz@@@carcfinal#1#2#3{%
  \pgf@process{#2}%
  \advance\pgf@x by \tikz@lastx
  \advance\pgf@y by \tikz@lasty
  \pgfpathmoveto{\pgfqpoint{\the\pgf@x}{\the\pgf@y}}%
  #1%
  \pgfpathmoveto{\pgfqpoint{\the\tikz@lastx}{\the\tikz@lasty}}%
  \let\tikz@@@arcfinal=\carc@orig@@@arcfinal
}

\def\tikz@carcfinal{%
  \tikz@lastxsaved=\tikz@lastx%
  \tikz@lastysaved=\tikz@lasty%
  \let\tikz@arcfinal=\carc@orig@arcfinal
  \tikz@scan@next@command%
}

\let\carc@orig@@@arcfinal=\tikz@@@arcfinal
\let\carc@orig@arcfinal=\tikz@arcfinal

\def\tikz@cchar{
    \pgfutil@ifnextchar i 
        {\tikz@circle}%
        {\pgfutil@ifnextchar h
            {\tikz@children}
            {\pgfutil@ifnextchar a 
                {\carc@call}
                \tikz@cochar
            }
        }
}%

\def\carc@call{\tikzset{centred arc}\tikz@scan@next@command}

\tikzset{
  centred arc/.code={%
    \let\tikz@@@arcfinal=\tikz@@@carcfinal
    \let\tikz@arcfinal=\tikz@carcfinal
  },
}
\makeatother

\title{\boldmath Four-point boundary connectivities in critical two-dimensional percolation  from conformal invariance}

\author[a,b]{Giacomo Gori}
\author[c]{Jacopo Viti}
\affiliation[a]{Dipartimento di Fisica e Astronomia ``Galileo Galilei'', Universit\`a di Padova, I-35131 Padova, Italy}
\affiliation[b]{CNR-IOM, Via Bonomea 265, 34136 Trieste, Italy}
\affiliation[c]{ECT \& International Institute of Physics, UFRN,  Campos Universit\' ario, Lagoa Nova  59078-970 Natal, Brazil}

\emailAdd{viti.jacopo@gmail.com}

\abstract
{We  conjecture an exact form for an universal ratio of  four-point cluster connectivities in the critical two-dimensional $Q$-color Potts model. We also provide analogous
results for the limit $Q\rightarrow 1$ that corresponds to percolation where the observable has a
logarithmic singularity. Our conjectures are tested against Monte Carlo simulations showing excellent agreement for $Q=1,2,3$.}

\begin{document} 
\maketitle
\flushbottom

\section{Introduction}
The study of the geometry of random two-dimensional fractals has revealed the emergence of a profound mathematical connection between probability
theory and stochastic processes~\cite{Schramm1, Schramm2, Schramm3, Smir} on one hand and quantum field theory  together with conformal
symmetry on the other~\cite{BB1, BB2, BB3, BB4, BB5, BB6, Cardy_SLE}. Historically, a number of exact results were derived for the fractal dimensions of
two-dimensional critical clusters in basic models of statistical mechanics such as percolation or Ising, built on the seminal
contribution~\cite{BPZ} and the so-called Coulomb Gas approach~\cite{CG}.  A deeper insight on how  conformal invariance  could be relevant to
describe geometrical observables followed  after~\cite{Cardy_perc} when  J. Cardy derived, with methods borrowed from (boundary) conformal
field theory~\cite{Cardy_surf, Cardy_verlinde}, an exact formula for the probability that at least a cluster should span the two horizontal sides 
of a rectangle in critical percolation~\cite{Langlands}. The success of this approach suggested that a geometrical problem, such as critical percolation, could be solvable in two dimensions due to the infinite-dimensional nature of the conformal group~\cite{BPZ}.

However, at the same time, it was noticed how the conformal algebra associated to geometrical phase transitions  could be more subtle~\cite{G}.
In particular as a conformal field theory, critical percolation should have vanishing central charge (denoted by $c$) since
its partition function does not depend on finite size effects~\cite{BCM1,BCM2}. However rightly at $c=0$ the stress-energy tensor is a null field and the
field theory if not trivial, cannot be unitary. Absence of unitarity has serious consequences on the Operator Product Expansion (OPE) and ultimately produces
logarithmic singularities in the four-point functions~\cite{RSal, Sal}.  Later, it was conjectured in~\cite{GL} that the OPE of two chiral fields with
scaling dimension $h\not=0$ at $c=0$ should have the following expansion ($z\in\mathbb C$)
\begin{equation}
\label{OPE1}
\lim_{z\rightarrow 0}\phi(z)\phi(0)=\frac{1}{z^{2h}}[1+\frac{2h}{b}z^2(t(0)+\log(z)T(0))+\dots],
\end{equation} 
where $T(z)$ is the null stress energy tensor and $t(z)$ was called its logarithmic partner. The parameter $b$ in Eq.~\eqref{OPE1}, termed the indecomposability parameter, is  a universal number characterizing the $c=0$ Conformal Field Theory (CFT)~\cite{GL, GanVa}. Its name, in particular, stems from the fact that
the fields $T$ and $t$ span a Jordan cell of dimension two which makes the conformal dilation operator non-diagonalizable.
CFTs
that are built upon indecomposable representations of the Virasoro algebra are called logarithmic~\cite{Ridout_rev}. 
They are supposed to be ubiquitous
in the study of random clusters and disordered two-dimensional systems~\cite{Cardy_stat, Cardy_revLog}. For detailed studies in higher dimensions, see also~\cite{Vi}.

During the last 
decade a lot of effort has been 
put in the classification of logarithmic CFTs 
with special success on finite domains;
see~\cite{Log_rev} and references therein. However
not many exact correlation functions have been explicitly calculated and tested in  statistical mechanics.
Important exceptions are G. Watts result~\cite{Watts} and other generalizations
of Cardy crossing formula on polygonal domains, such as hexagons or octagons~\cite{Flores1, Flores2, Flores3, Flores4, FKJS}.
In particular, logarithmic singularities in crossing probabilities are hidden into higher-point correlation functions~\cite{Simmons, FKJS};
for instance the six-point functions of the field $\phi_{1,2}$ in the notations of Eq.~\eqref{scaling_b}. Such a  field has vanishing scaling dimensions
at $c=0$
and its four-point function cannot be logarithmic~\cite{Cardy_perc}, cf. Eq.~\eqref{OPE1}. 
We should also mention  that the study of logarithmic conformal field theories in the bulk is considerably harder than on a finite geometry,
due to the constraints of crossing symmetry. Recent developments for three~\cite{DV, Liouville, IJS} and four-point functions~\cite{PRS},
are based on a conformal bootstrap approach to Liouville theory for $c<1$.

In this paper we complement  those existing results, introducing and exactly determining  a geometrical
observable that explicitly shows logarithmic behavior at criticality.  We focus on the $Q$-color Potts model~\cite{Wu} on a bounded domain
and construct a  ratio between four-point cluster connectivities, see Fig.~\ref{fig_conn}. We then follow closely~\cite{Cardy_perc} and
symmetry arguments to obtain a fully analytic expression for such a quantity in terms of Virasoro conformal blocks~\cite{BPZ}. 
The main  technical assumption of our approach is that the functions in Eq.~\eqref{ratio_r} solve a  third order differential equation
that is associated to a null vector for a field with non-zero scaling dimension.  Null vector decoupling for a non-unitary CFT is not granted \textit{a priori} and requires care. However for the specific case  considered here, it was proven self-consistent~\cite{GL} at $c=0$ and led to the conjecture  $b=-5/8$ in Eq.~\eqref{OPE1} for boundary percolation. The same value for $b$ was later re-derived in~\cite{DJS} by algebraic means and it is now accepted as  a universal number characterizing this universality class~\cite{Cardy_revLog}. Our results in Sec.~\ref{sec3} for boundary percolation could be then considered  both a non-trivial application and a further test of the ideas in~\cite{GL}; in particular the OPE~\eqref{OPE1}. 
As noticed in~\cite{Cardy_revLog, VJS}, our geometrical observable is also logarithmic at $Q=2$, i.e. in the (extended~\cite{PRZ}) Ising model.
The latter was already analyzed in~\cite{GV17} by the
same authors but we recast it here in a more general context. We also provide numerical checks of all our results through high-precision
Monte Carlo simulations and further
extend the analysis in~\cite{GV17} to the three-color and four-color Potts model.  These cases, although conceptually analogous and technically simpler than $Q=1$ and $Q=2$ have been not considered before.

The outline of the rest of the paper is as follows. In Sec.~\ref{sec1} we introduce the $Q$-color Potts model and the geometrical observable $R$.
In Sec.~\ref{sec2} we show how this ratio of cluster connectivities can be obtained from conformal invariance and derive explicit analytic
expressions in Sec.~\ref{sec3}. The comparison of the CFT predictions against Monte Carlo simulations
is addressed in Sec.~\ref{sec4}. After the conclusions, three technical appendices complete the paper.

\section{Four-point boundary connectivities in the $Q$-color Potts model}
\label{sec1}
\begin{figure}[t]
\centering
\begin{tikzpicture}[scale=0.8]
\draw[thick] (0cm,0cm) circle(1.8cm);
\draw[blue, very thick](1.8/2, 1.8*1.713/2)--(-1.8/2, 1.8*1.713/2);
\draw[blue, very thick](1.8/2, -1.8*1.713/2)--(-1.8/2, -1.8*1.713/2);
\node[red] at  (1.8/2, 1.8*1.713/2) {\Large{$\bullet$}};
\node[red] at  (-1.8/2, 1.8*1.713/2) {\Large{$\bullet$}};
\node[red] at  (-1.8/2, -1.8*1.713/2) {\Large{$\bullet$}};
\node[red] at  (1.8/2, -1.8*1.713/2) {\Large{$\bullet$}};
\node[above] at  (1.8/2+.07, 1.8*1.713/2+.07) {\large{$x_2$}};
\node[above] at  (-1.8/2-.035, 1.8*1.713/2+.07) {\large{$x_1$}};
\node[below] at  (-1.8/2-.035, -1.8*1.713/2-.07) {\large{$x_4$}};
\node[below] at  (1.8/2+.07, -1.8*1.713/2-.07) {\large{$x_3$}};
\node at (0,2.5){$P_{(12)(34)}$};
\node at (0,0){\Large{$\mathcal{D}$}};
\begin{scope}[xshift=5cm]
 \draw[thick] (0cm,0cm) circle(1.8cm);
 \draw[blue, very thick](1.8/2, 1.8*1.713/2)--(1.8/2, -1.8*1.713/2);
\draw[blue, very thick](-1.8/2, 1.8*1.713/2)--(-1.8/2, -1.8*1.713/2);
\node[red] at  (1.8/2, 1.8*1.713/2) {\Large{$\bullet$}};
\node[red] at  (-1.8/2, 1.8*1.713/2) {\Large{$\bullet$}};
\node[red] at  (-1.8/2, -1.8*1.713/2) {\Large{$\bullet$}};
\node[red] at  (1.8/2, -1.8*1.713/2) {\Large{$\bullet$}};
\node[above] at  (1.8/2+.07, 1.8*1.713/2+.07) {\large{$x_2$}};
\node[above] at  (-1.8/2-.035, 1.8*1.713/2+.07) {\large{$x_1$}};
\node[below] at  (-1.8/2-.035, -1.8*1.713/2-.07) {\large{$x_4$}};
\node[below] at  (1.8/2+.07, -1.8*1.713/2-.07) {\large{$x_3$}};
\node at (0,2.5){$P_{(14)(23)}$};
\node at (0,0){\Large{$\mathcal{D}$}};
\end{scope}
\begin{scope}[xshift=10cm]
 \draw[thick] (0cm,0cm) circle(1.8cm);
\draw[blue, very thick](1.8/2, 1.8*1.713/2)--(-1.8/2, -1.8*1.713/2);
\draw[blue, very thick](-1.8/2, 1.8*1.713/2)--(1.8/2, -1.8*1.713/2);
\node[red] at  (1.8/2, 1.8*1.713/2) {\Large{$\bullet$}};
\node[red] at  (-1.8/2, 1.8*1.713/2) {\Large{$\bullet$}};
\node[red] at  (-1.8/2, -1.8*1.713/2) {\Large{$\bullet$}};
\node[red] at  (1.8/2, -1.8*1.713/2) {\Large{$\bullet$}};
\node[above] at  (1.8/2+.07, 1.8*1.713/2+.07) {\large{$x_2$}};
\node[above] at  (-1.8/2-.035, 1.8*1.713/2+.07) {\large{$x_1$}};
\node[below] at  (-1.8/2-.035, -1.8*1.713/2-.07) {\large{$x_4$}};
\node[below] at  (1.8/2+.07, -1.8*1.713/2-.07) {\large{$x_3$}};
\node at (0,2.5){$P_{(1234)}$};
\node[left] at (0,0){\Large{$\mathcal{D}$}};
\end{scope}
\end{tikzpicture}
\caption{Four points $x_1, x_2, x_3$ and $x_4$ are marked on the boundary of a simply connected domain $\mathcal{D}$ embedded into a regular two-dimensional lattice. The function $P_{(12)(34)}$ is the probability that $x_1$ and $x_2$ are connected into one FK cluster while $x_3$ and $x_4$ are connected into a different FK cluster. Analogously $P_{(14)(23)}$ is the probability that $x_1$ and $x_4$   are connected into one FK cluster while $x_2$ and $x_3$ are connected into a different FK cluster. Finally $P_{(1234)}$ is the probability that all the points belong to the same FK cluster. The non-normalized probability measure of any configurations
(i.e. of any random graphs $\mathcal{G}$) is given in Eq.~(\ref{pf_random}).}
\label{fig_conn}
\end{figure}
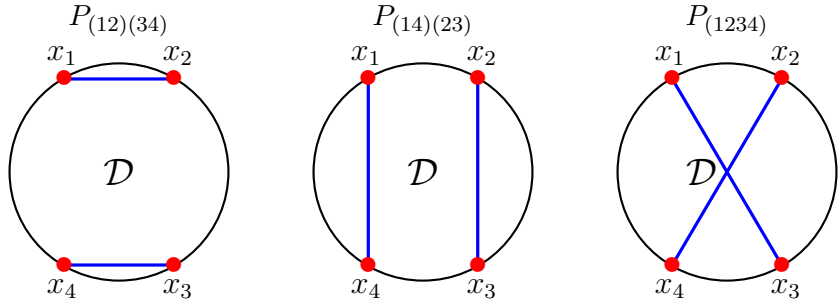
The $Q$-color Potts model~\cite{Wu} is defined by the Hamiltonian $(J>0)$
\begin{equation}
\label{Potts}
H_{Q}=-J\sum_{\langle x,y\rangle}\delta_{s(x),s(y)},~~~s(x)=1,\dots, Q 
\end{equation}
where the spin variable  $s(x)$ takes only positive integer values  up to $Q$, and the sum extends over next-neighboring sites on
a certain bounded domain $\mathcal{D}$ embedded into a two-dimensional  regular lattice. The boundary conditions for the spin are free.
The Potts partition function $Z(Q)=\sum_{\{s(x)\}}e^{-H_{Q}}$ admits a well known graph expansion, the so-called Fortuin and Kasteleyn~\cite{FK1,FK2} representation.
Let $p=1-e^{-J}$ the probability of drawing a bond between two next neighboring lattice sites in $\mathcal D$, then it turns out, up to a multiplicative constant,
\begin{equation}
\label{pf_random}
Z(Q)=\sum_{\mathcal{G}}p^{n_b}(1-p)^{\bar{n}_b}Q^{N_c}.
\end{equation} 
In Eq.~(\ref{pf_random}) above,  $n_b$ (resp. $\bar{n}_b$) is the number of occupied (resp. empty) bonds in the domain $\mathcal{D}$.
Connected components, including isolated points, inside a graph $\mathcal{G}$ are called clusters (FK clusters). 
Each graph contains $N_c$ clusters  in which the Potts spins are
forced to have the same  color, 
hence the factor $Q^{N_c}$.
When $Q=2$, Eq.~(\ref{pf_random}) is the high-temperature expansion
of the Ising model. 
Although the partition function $Z(Q)$ can be defined for any complex
$Q$, in this paper we will consider only positive integer values including however  $Q=1$ that corresponds
to the percolation problem. In particular, we will be interested in determining boundary
connectivities in the $Q$-color Potts model.

Connectivities in the $Q$-color Potts model are probabilities
that a certain set of points marked by $x_1,\dots,x_n$ are partitioned
into FK clusters. A non-normalized probability measure for the allowed graph configurations is given by $m(\mathcal{G})=p^{n_b}(1-p)^{\bar{n}_b}Q^{N_c}$ according to Eq.~(\ref{pf_random}). The normalized probability measure for the graphs  would be of course $Z^{-1}m(\mathcal{G})$ as $Z(Q)=1$ only at $Q=1$; the normalization factor is however not essential here since only ratios of connectivities will be considered. Moreover, if we  focus on configurations in which
$n$ points are on the boundary, it
can be shown~\cite{DVPotts} that  the number of linearly independent
connectivities is also the number of non-crossing non-singleton partitions of a set of $n$ elements
(also known as Riordan numbers~\cite{Bernhard}).  In geometrical terms, linear independent boundary connectivities correspond indeed~\cite{DVPotts} to configurations where none of the boundary points belongs  to an isolated FK cluster (i.e. it is disconnected from all the other points).  In particular
if $n=4$ there are only three linearly independent
four-point boundary connectivities that are schematically represented
in Fig.~\ref{fig_conn}. With obvious notations
such four-point functions will be denoted by $P_{(12)(34)}$, $P_{(14)(23)}$ and $P_{(1234)}$, see again Fig.~\ref{fig_conn}. Notice that, due to the planarity of the domain $\mathcal{D}$, we have $P_{(13)(24)}=0$. This important simplification does not occur  when the four-points are in the bulk; in such a case the number of linear independent four-point connectivities is four.

For $1\leq Q\leq 4$, the Potts model undergoes a second order ferromagnetic phase transition for a critical reduced inverse temperature $J=J_c$.
The ferromagnetic phase transition is instead of the first order for $Q>4$.
In geometrical terms, at $J>J_c$ ($p>p_c$) there is a finite probability that any bulk point will be connected to the boundary of $\mathcal{D}$. Such a
probability vanishes as a power law as $J\rightarrow J_c^+$ with a critical exponent that coincides with the one of the one-point function of the
order  parameter; for instance~\cite{Grimmet}.

At $J=J_c$, and in the scaling limit when the mesh of the lattice  is sent to zero, connectivities, although strictly speaking vanishing, are conjectured to be conformally covariant~\cite{Langlands}. It is then useful to define the dimensionless conformal invariant ratio
\begin{equation}
\label{ratio_r}
R=\frac{P_{(14)(23)}}{P_{(14)(23)}+P_{(12)(34)}+P_{(1234)}}.
\end{equation}
 Let us aso clarify the statement that $R$ is a conformal invariant quantity. If $x_j\equiv(s_j,t_j)$ belongs to $\mathcal{D}$  we introduce
complex coordinates $w_j=s_j+it_j\in\mathbb C$. Then the bounded domain $\mathcal{D}$ (for instance the unit disk) can be mapped conformally
through the mapping $z=z(w)$ into the upper half plane. When the points $w_j$ are on the boundary of $\mathcal{D}$, they are mapped on
the real axis and we
can always choose $z_1<z_2<z_3<z_4$. Conformal invariance implies that $R$ calculated on the upper half plane  is only a function of the
anharmonic ratio
\begin{equation}
\label{eta}
\eta=\frac{z_{21}z_{43}}{z_{31}z_{42}}
\end{equation}
where $z_{ij}=z_i-z_j$. We can then determine $R$ on
the original domain $\mathcal{D}$  simply replacing $z_j=z(w_j)$ ($j=1,\dots, 4$) into the expression for $R(\eta)$.
In the next sections, we will obtain exactly $R$. The conjecture for Eq.~\eqref{ratio_r}   is valid
for all the integer values $Q=1,2,3,4$ and will be eventually tested 
against Monte Carlo simulations.
Anticipating the content of the remaining sections,  the  reader will find a plot of our theoretical predictions for the universal ratio in Eq.~\eqref{ratio_r} in Fig.~\ref{figrs}.

It is important to observe that, although their leading short-distance singularities are the same, four-point boundary connectivities cannot be obtained
from the knowledge of the boundary correlation functions of the Potts order parameter. A paradigmatic example~\cite{DVPotts} is $Q=2$, where the unique  boundary four-point function of the spin
is a linear combination of the three connectivities in Fig.~\ref{fig_conn}.
 Three-point boundary connectivities for  the $Q$-color Potts model (and in particular at $Q=3$) could  be determined instead  from the known solutions~\cite{Runa, Rund} of the boundary bootstrap equations for the Minimal Moldels.

\section{Duality and conformal symmetry}
\label{sec2}
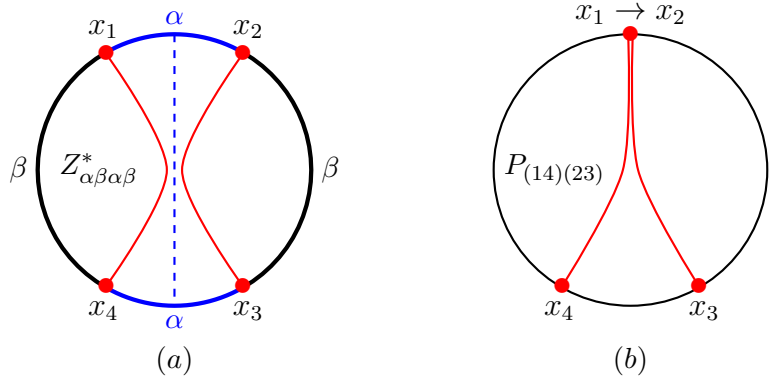
\begin{figure}[t]
 \centering
\begin{tikzpicture}
\draw[thick] (0cm,0cm) circle(1.8cm);
\draw [blue, ultra thick, domain=60:120] plot ({1.8*cos(\x)}, {1.8*sin(\x)});
\draw [ ultra thick, domain=120:240] plot ({1.8*cos(\x)}, {1.8*sin(\x)});
\draw [blue, ultra thick,  domain=240:300] plot ({1.8*cos(\x)}, {1.8*sin(\x)});
\draw [ultra thick, domain=-60:60] plot ({1.8*cos(\x)}, {1.8*sin(\x)});
\node[red] at  (1.8/2, 1.8*1.713/2) {\Large{$\bullet$}};
\node[red] at  (-1.8/2, 1.8*1.713/2) {\Large{$\bullet$}};
\node[red] at  (-1.8/2, -1.8*1.713/2) {\Large{$\bullet$}};
\node[red] at  (1.8/2, -1.8*1.713/2) {\Large{$\bullet$}};



\draw[red, thick] plot [smooth] coordinates {(-1.8/2, 1.8*1.713/2) (-0.1,0)
(-1.8/2, -1.8*1.713/2)};
\draw[red, thick] plot [smooth] coordinates {(1.8/2, 1.8*1.713/2) (0.1,0)
(1.8/2, -1.8*1.713/2)};
\draw[dashed, thick, blue](0,1.8)--(0,-1.8);

\node[above] at  (1.8/2+.07, 1.8*1.713/2+.07) {\large{$x_2$}};
\node[above] at  (-1.8/2-.035, 1.8*1.713/2+.07) {\large{$x_1$}};
\node[below] at  (-1.8/2-.035, -1.8*1.713/2-.07) {\large{$x_4$}};
\node[below] at  (1.8/2+.07, -1.8*1.713/2-.07) {\large{$x_3$}};

\node[blue, above] at (0,1.8) {$\alpha$};
\node[right] at (1.8,0) {$\beta$};
\node[below, blue] at (0,-1.8) {$\alpha$};
\node[left] at (-1.8,0) {$\beta$};
\node at (-1,0) {$Z^*_{\alpha\beta\alpha\beta}$};
\node[below] at (0,-2.2) {$(a)$};
\begin{scope}[xshift=6cm]
\draw[red, thick] plot [smooth] coordinates {(-0.03, 1.8) (-0.1,0)
(-1.8/2, -1.8*1.713/2)};
\draw[red, thick] plot [smooth] coordinates {(0.03, 1.8) (0.1,0)
(1.8/2, -1.8*1.713/2)};
\node[above] at  (0, 1.8) {\large{$x_1\rightarrow x_2$}};
\node[below] at  (-1.8/2-.035, -1.8*1.713/2-.07) {\large{$x_4$}};
\node[below] at  (1.8/2+.07, -1.8*1.713/2-.07) {\large{$x_3$}};
\draw[thick] (0cm,0cm) circle(1.8cm);
\node[red] at  (0, 1.8) {\Large{$\bullet$}};
\node[red] at  (-1.8/2, -1.8*1.713/2) {\Large{$\bullet$}};
\node[red] at  (1.8/2, -1.8*1.713/2) {\Large{$\bullet$}};
\node at (-1,0) {$P_{(14)(23)}$};
\node[below] at (0,-2.2) {$(b)$};
\end{scope}
\end{tikzpicture}
\caption{(a) Partition function $Z^*_{\alpha\beta\alpha\beta}$. It is drawn a dual FK cluster (dashed line) leading to a vertical crossing.
These cluster configurations are dual to the ones contributing to $P_{(14)(23)}$. In particular it is not possible to connect $x_1$ with $x_3$ or $x_2$ with $x_4$ without crossing the dual dashed cluster. (b) The limit $x_1\rightarrow x_2$ in the function $P_{(14)(23)}$ produces
configurations where two distinct FK clusters meet at $x_2$. Field theoretically this is interpreted as the insertion of the field $\phi_{1,5}$ at the boundary point $x_2$.}
\label{fig_dual}
\end{figure}

\textit{Boundary-condition-changing operators and duality---}The quantum field theory that describes the critical large-distance fluctuations
of the two-dimensional $Q$-color Potts model is a conformal field theory (CFT)~\cite{DF} with central charge
\begin{equation}
\label{cc}
c=1-\frac{6}{p(p+1)}
\end{equation} 
and the parameter $Q=4\cos^2[\pi/(p+1)]$. The central charge in Eq.~(\ref{cc}) enters the commutation relations of the Virasoro algebra generators
\begin{equation}
\label{Vir}
[L_{n},L_{m}]=(n-m)L_{n+m}+\frac{c}{12}\delta_{n+m,0}n(n^2-1).
\end{equation}
In principle $p\in\mathbb R$, however since $Q$ is integer and $1\leq Q\leq 4$, we only consider the cases $p=2$ (percolation), $p=3$ (Ising), $p=5$ (three-color Potts model) and $p\rightarrow\infty$ (four-color Potts model). Notice that at $Q=1$ the central charge in Eq.~(\ref{cc}) is zero, a well-known circumstance of critical percolation that in particular implies that in such a case the CFT is non-unitary (the only unitary CFT with zero central charge is indeed trivial). This lack of unitary reflects itself into the presence, as we will discuss,
of logarithmic singularities~\cite{G, RSal, Sal, GL} in the four-point connectivities. Let us now briefly review some basics of (boundary) CFTs in two dimensions~\cite{Cardy_surf, Cardy_verlinde}. 
When a scaling field is inserted at the boundary of a planar  domain $\mathcal{D}$, its scaling dimensions are eigenvalues of the operator $L_0$~\cite{Cardy_surf}.  The scaling dimensions of a primary field (for a definition see~\cite{Cardy_book})
$\phi_{r,s}$ sitting at the boundary are then given by
\begin{equation}
\label{scaling_b}
h_{r,s}=\frac{[r(p+1)-sp]^2-1}{4p(p+1)};
\end{equation}
and it turns out that for our purposes $r,s$ are positive integers. In such a case the fields $\phi_{r,s}$ are also dubbed degenerate~\cite{BPZ} and their correlation functions satisfy partial differential equations of degree $rs$. The operator content of the CFT depends on the boundary conditions.
For the $Q$-color Potts model natural boundary conditions for the spin variable on $\mathcal{D}$ are either free or fixed to a definite color $\alpha=1,\dots,Q$. Remarkably, conformal symmetry is however also compatible with inhomogeneous boundary conditions that are associated to the insertion of scaling fields at the boundary called boundary-condition-changing (bcc) operators~\cite{Cardy_verlinde}. For instance if the value of the spin on the boundary switches from  $\alpha$  to  $\beta\not=\alpha$ nearby  $x$, such a discontinuity in the boundary conditions  is realized, in the scaling limit, by the insertion of a scaling field $\phi_{\alpha\beta}(x)$. In particular, the
bcc operator $\phi_{\alpha\beta}$ can be identified with a field $\phi_{r,s}$, whose scaling dimensions are given in Eq.~(\ref{scaling_b}); we will discuss which field in the next subsection. Before we remind, as pointed out first in~\cite{Cardy_perc}, how correlation functions of
bcc operators can be related to connectivities in the $Q$-color Potts model. The argument exploits the duality transformation of the Potts partition function on a planar domain $\mathcal{D}$; in particular the mapping also requires a transformation of the lattice.
\begin{figure}[t]
\begin{tikzpicture}[scale=0.8]
\draw[blue, ultra  thick] (0,0) carc(60:120:1.8);
\draw[blue, ultra  thick] (0,0) carc(240:300:1.8);
\draw[ultra  thick] (0,0) carc(120:240:1.8);
\draw[ultra  thick] (0,0) carc(300:360:1.8);
\draw[ultra  thick] (0,0) carc(0:60:1.8);
\node[red] at  (1.8/2, 1.8*1.713/2) {\Large{$\bullet$}};
\node[red] at  (-1.8/2, 1.8*1.713/2) {\Large{$\bullet$}};
\node[red] at  (-1.8/2, -1.8*1.713/2) {\Large{$\bullet$}};
\node[red] at  (1.8/2, -1.8*1.713/2) {\Large{$\bullet$}};
\node[above] at  (-1.8/2-.035, 1.8*1.713/2+.07) {\large{$x_1$}};
\node[above] at  (1.8/2-.035, 1.8*1.713/2+.07) {\large{$x_2$}};
\node[below] at  (-1.8/2-.035, -1.8*1.713/2-.07) {\large{$x_4$}};
\node[below] at  (1.8/2+.07, -1.8*1.713/2-.07) {\large{$x_3$}};
\node[above] at (0,1.8) {$\alpha$};
\node[right] at (1.8,0) {$\beta$};
\node[below] at (0,-1.8) {$\alpha$};
\node[left] at (-1.8,0) {$\beta$};
\node at (0,0){\Large{$\mathcal{D}$}};
\begin{scope}[xshift= 5cm]
\draw[blue, ultra  thick] (0,0) carc(60:120:1.8);
\draw[ultra  thick] (0,0) carc(300:360:1.8);
\draw[ultra  thick] (0,0) carc(0:60:1.8);
\draw[yellow!80!red, ultra  thick] (0,0) carc(240:300:1.8);
\draw[green, ultra  thick] (0,0) carc(120:240:1.8);
\node[red] at  (1.8/2, 1.8*1.713/2) {\Large{$\bullet$}};
\node[red] at  (-1.8/2, 1.8*1.713/2) {\Large{$\bullet$}};
\node[red] at  (-1.8/2, -1.8*1.713/2) {\Large{$\bullet$}};
\node[red] at  (1.8/2, -1.8*1.713/2) {\Large{$\bullet$}};
\node[above] at  (-1.8/2-.035, 1.8*1.713/2+.07) {\large{$x_1$}};
\node[above] at  (1.8/2-.035, 1.8*1.713/2+.07) {\large{$x_2$}};
\node[below] at  (-1.8/2-.035, -1.8*1.713/2-.07) {\large{$x_4$}};
\node[below] at  (1.8/2+.07, -1.8*1.713/2-.07) {\large{$x_3$}};
\node[above] at (0,1.8) {$\alpha$};
\node[right] at (1.8,0) {$\beta$};
\node[below] at (0,-1.8) {$\gamma$};
\node[left] at (-1.8,0) {$\delta$};
\node at (0,0){\Large{$\mathcal{D}$}};
\end{scope}
\begin{scope}[xshift= 10cm]
\draw[blue, ultra  thick] (0,0) carc(60:120:1.8);
\draw[ultra  thick] (0,0) carc(300:360:1.8);
\draw[ultra  thick] (0,0) carc(0:60:1.8);
\draw[blue, ultra  thick] (0,0) carc(240:300:1.8);
\draw[yellow!80!red, ultra  thick] (0,0) carc(120:240:1.8);
\node[red] at  (1.8/2, 1.8*1.713/2) {\Large{$\bullet$}};
\node[red] at  (-1.8/2, 1.8*1.713/2) {\Large{$\bullet$}};
\node[red] at  (-1.8/2, -1.8*1.713/2) {\Large{$\bullet$}};
\node[red] at  (1.8/2, -1.8*1.713/2) {\Large{$\bullet$}};
\node[above] at  (-1.8/2-.035, 1.8*1.713/2+.07) {\large{$x_1$}};
\node[above] at  (1.8/2-.035, 1.8*1.713/2+.07) {\large{$x_2$}};
\node[below] at  (-1.8/2-.035, -1.8*1.713/2-.07) {\large{$x_4$}};
\node[below] at  (1.8/2+.07, -1.8*1.713/2-.07) {\large{$x_3$}};
\node[above] at (0,1.8) {$\alpha$};
\node[right] at (1.8,0) {$\beta$};
\node[below] at (0,-1.8) {$\alpha$};
\node[left] at (-1.8,0) {$\gamma$};
\node at (0,0){\Large{$\mathcal{D}$}};
\end{scope}
\begin{scope}[xshift= 15cm]
\draw[blue, ultra  thick] (0,0) carc(60:120:1.8);
\draw[ultra  thick] (0,0) carc(300:360:1.8);
\draw[ultra  thick] (0,0) carc(0:60:1.8);
\draw[yellow!80!red, ultra  thick] (0,0) carc(240:300:1.8);
\draw[ultra  thick] (0,0) carc(120:240:1.8);
\node[red] at  (1.8/2, 1.8*1.713/2) {\Large{$\bullet$}};
\node[red] at  (-1.8/2, 1.8*1.713/2) {\Large{$\bullet$}};
\node[red] at  (-1.8/2, -1.8*1.713/2) {\Large{$\bullet$}};
\node[red] at  (1.8/2, -1.8*1.713/2) {\Large{$\bullet$}};
\node[above] at  (-1.8/2-.035, 1.8*1.713/2+.07) {\large{$x_1$}};
\node[above] at  (1.8/2-.035, 1.8*1.713/2+.07) {\large{$x_2$}};
\node[below] at  (-1.8/2-.035, -1.8*1.713/2-.07) {\large{$x_4$}};
\node[below] at  (1.8/2+.07, -1.8*1.713/2-.07) {\large{$x_3$}};
\node[above] at (0,1.8) {$\alpha$};
\node[right] at (1.8,0) {$\beta$};
\node[below] at (0,-1.8) {$\gamma$};
\node[left] at (-1.8,0) {$\beta$};
\node at (0,0){\Large{$\mathcal{D}$}};
\end{scope}
\end{tikzpicture}
\caption{Four boundary partition functions (from left to right) $Z^*_{\alpha\beta\alpha\beta}, Z^*_{\alpha\beta\gamma\delta},~Z^*_{\alpha\beta\alpha\gamma}$ and $Z^*_{\alpha\beta\gamma\alpha}$.  Different colors correspond to different fixed boundary conditions for the (dual) Potts spin. The following relation holds $Z^{*}_{\alpha\beta\alpha\beta}+Z^*_{\alpha\beta\gamma\delta}=Z^{*}_{\alpha\beta\alpha\gamma}+Z^{*}_{\alpha\beta\gamma\alpha}$, showing that only three of them are linearly independent. They are related~\cite{DVPotts, WY} through a duality transformation to the three linearly independent connectivities in Fig.~\ref{fig_conn}, calculated with free boundary conditions. At the critical point  and in the scaling limit, they are also proportional, to the four-point functions of the bcc $\phi_{\alpha\beta}$.}
\label{fig_dual_part}
\end{figure}
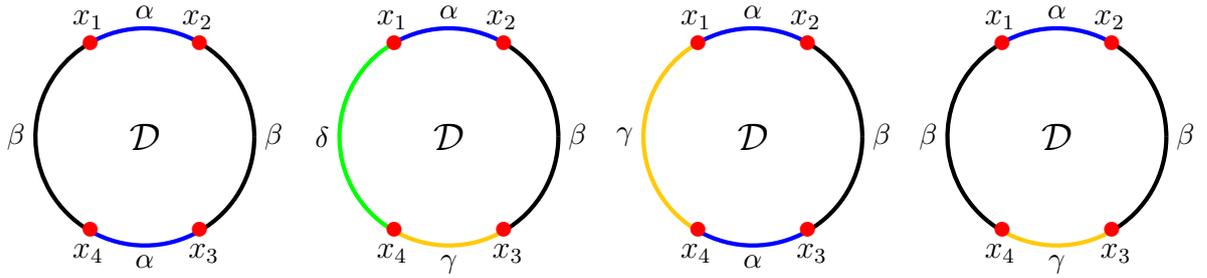
A duality transformation~\cite{DVPotts} relates   partition functions on the dual lattice  with fixed boundary conditions and in the ordered phase $(J^*\geq J_c)$ to connectivities
calculated with free boundary conditions on the original lattice in the disordered phase $(J\leq J_c)$.For a definition of the dual of an FK cluster and the lattice-dependent mapping $J^*(J)$, we refer to Sec.II of~\cite{Wu}. There are four different dual partition functions, involving four boundary points, that we could consider for $J^*\geq J_c$. They are denoted by $Z^*_{\alpha\beta\alpha\beta}, ~Z^*_{\alpha\beta\gamma\delta},~Z^*_{\alpha\beta\alpha\gamma}$ and $Z^*_{\alpha\beta\gamma\beta}$ and  represented in Fig.~\ref{fig_dual_part}. Regions coloured differently at the boundary of $\mathcal{D}$ correspond to regions with different fixed boundary conditions for the dual spins. It is also important to remark that the following relation~\cite{DVPotts, WY} holds: $Z^*_{\alpha\beta\alpha\beta}+Z^*_{\alpha\beta\gamma\delta}=Z^*_{\alpha\beta\alpha\gamma}+Z^*_{\alpha\beta\gamma\beta}$, showing that actually only three of them are linearly independent. 
Among the duality relations~\cite{WY, DVPotts}, involving the four-partition functions above and the three linearly independent connectivities with free boundary conditions in Fig.~\ref{fig_conn}, we will only need (see the discussion below Eq.~\eqref{OPE})  the following
\begin{equation}
\label{duality}
\left. Z^{*}_{\alpha\beta\alpha\beta}(x_1,x_2,x_3,x_4)\right|_{J^*\geq J_c}=A'\left(P_{(12)(34)}+P_{(14)(23)}+P_{(1234)})\right|_{J\leq J_c}.
\end{equation}
In Eq.~\eqref{duality}, $A'$ is a  lattice-dependent normalization constant that however does not depend on $x_1,\dots, x_4$. Eq.~(\ref{duality}) was obtained in~\cite{DVPotts} directly in the scaling limit, but  it can also be understood as follows. Perform an FK graph expansion of the partition function $Z^*_{\alpha\beta\alpha\beta}$, then the dual FK clusters cannot connect regions where the spins are fixed to have different colors at the boundary. We can distinguish three cases:
\begin{itemize}
\item Dual graph configurations contain at least a dual cluster connecting the two regions  with boundary conditions $\beta$ (there is an horizontal dual crossing). Applying a duality transformation these configurations are in one-to-one correspondence with the ones that contribute to $P_{(12)(34)}$.
\item Dual graph configurations contain at least a dual cluster connecting the two regions  with boundary conditions $\alpha$ (there is a vertical dual crossing). Applying a duality transformation these configurations are in one-to-one correspondence with the ones that contribute to $P_{(14)(23)}$, see Fig.~\ref{fig_dual}a.
\item Dual graph configurations do not contain any cluster that connects regions on the boundary with the same color (there are no dual crossings). Applying a duality transformation these configurations are in one-to-one correspondence with $P_{(1234)}$. 
\end{itemize}
Notice that cannot be simultaneous horizontal and vertical crossings; this possibility was instead investigated in~\cite{Watts}. Summing over the three possibilities we obtain Eq.~(\ref{duality}).
The partition function $Z^{*}_{\alpha\beta\alpha\beta}(x_1,x_2,x_3,x_4)$ is in turn proportional~\cite{Cardy_verlinde, Cardy_perc} in the scaling limit to the four-point function $\langle\phi_{\alpha\beta}(x_1)\phi_{\beta\alpha}(x_2)\phi_{\alpha\beta}(x_3)\phi_{\beta\alpha}(x_4)\rangle$.  At the critical self-dual point ($J=J^*=J_c$) we then conclude that 
\begin{equation}
\label{bcc_conn}
\langle\phi_{\alpha\beta}(x_1)\phi_{\beta\alpha}(x_2)\phi_{\alpha\beta}(x_3)\phi_{\beta\alpha}(x_4)\rangle=A''P_t(x_1,x_2,x_3,x_4),
\end{equation}
with $A''$ constant and $P_t\equiv P_{(12)(34)}+P_{(14)(23)}+P_{(1234)}$.  Eq.~\eqref{bcc_conn} will be used  to derive the expansion in conformal blocks in Eq.~(\ref{ratiofin}) for the universal ratio $R$ in Eq.~\eqref{ratio_r}.

\noindent
\textit{Conformal blocks and the universal ratio $R$---} Following~\cite{Cardy_perc}, we identify the bcc operator $\phi_{\alpha\beta}$ with the field $\phi_{1,3}$, whose scaling dimensions are given (cf. Eq.~(\ref{scaling_b})) by
\begin{equation}
\label{dim}
h(p)=\frac{p-1}{p+1}.
\end{equation} 
The identification holds for any integer $1\leq Q\leq 4$.
Let us now consider the boundary four-point function of the field $\phi_{1,3}$; as discussed in Sec.~\ref{sec2} we work on the upper half plane, denoted hereafter by $\mathbb H$. The four points are then ordered on the real axis and chosen such that $z_1<z_2<z_3<z_4$. Exploiting global conformal symmetry on the upper half plane (i.e. $SL(2,\mathbb R)$  Moebius transformations), the four-point function of $\phi_{1,3}$ can be written as
\begin{equation}
\label{corr_13}
\langle\phi_{1,3}(z_1)\phi_{1,3}(z_2)\phi_{1,3}(z_3)
\phi_{1,3}(z_4)\rangle_{\mathbb H}=
\frac{1}{(z_{12}z_{34})^{2h}}\frac{G(\eta)}{(1-\eta)^{2h}}.
\end{equation}
The function $G(\eta)$ solves~\cite{BPZ} an Ordinary Differential Equation (ODE) of degree $3$ that can be obtained
by the condition of decoupling of the null-vector at level three in the Verma module of $\phi_{1,3}$.  Although null vector decoupling for a non-unitary CFT requires care, for the specific case of operators on the first column of the Kac table, this appears consistent with all available predictions based on chiral logarithmic CFTs~\cite{GL, GanVa}. It has been also employed in~\cite{KZ} for an analogous geometrical problem in percolation. The derivation of such a differential equation
is standard, the  reader can consult for instance~\cite{DMS}. It turns out
\begin{multline}
\label{ode}
 6 (1-h) h^2 (-1+2 \eta) G(\eta) +\left[2 (-1+\eta) \eta-3 h\left(1-5 \eta+5 \eta^2\right)+h^2 \left(3-19 \eta+19 \eta^2
 \right)\right] G'(\eta)\\
  +(-1+\eta) \eta \left[(-2+4 h+4 \eta-8 h \eta)G''(\eta)+(-1+\eta) \eta G'''(\eta)\right]=0,
\end{multline}
and $h$ depends on $p$ as in Eq.~\eqref{dim} whereas $p$ is related to the central charge and $Q$ by Eq.~\eqref{cc}. The equation above
is of Fuchsian type with regular singular points in $\eta=0,1$ and $\infty$, therefore we can write its Frobenius series near
$\eta=0$ as $G_{\rho}(\eta)=\eta^{\rho}\sum_{k=0}^{\infty} a_k\eta^k$ where conventionally we set $a_0=1$. The series has radius of convergence
$|\eta|< 1$ however this is enough for our purposes since by a global conformal transformation we can set $z_1=0$, $z_2=\eta$,  $z_3=1$ and $z_4=\infty$,
and therefore recalling that the four points are ordered on the boundary $\eta\in[0,1]$. Notice also that Eq.~\eqref{ode} is symmetric under the
transformation $\eta\rightarrow(1-\eta)$. The exponent $\rho$ solves the indicial equation
\begin{equation}
\label{inde}
\rho(\rho-h)(\rho-3h-1)=0
\end{equation}
and the three roots
coincide with the scaling dimensions $h_{1,1}$, $h_{1,3}$ and $h_{1,5}$ given in Eq.~\eqref{scaling_b}. This is of course expected since~\cite{BPZ}
the roots of Eq.~\eqref{inde} are the scaling dimensions of the leading singularities produced
in the OPE $\lim_{z_1\rightarrow z_2}\phi_{1,3}(z_1)\phi_{1,3}(z_2)$ as it can be seen from Eq.~\eqref{corr_13}; in particular $h_{1,5}=3h+1$. 
When the differences between the roots of the indicial equation are not integer and always for the largest root, the Frobenius series are directly
the Virasoro
conformal blocks of the CFT. Denoting by $F^c_{\rho}$ the four-point conformal block where we fixed the external legs to be the
field $\phi_{1,3}$ (of dimension $h(c)$) and the internal field has scaling dimension $\rho$, we have
\begin{equation}
\label{conf}
 F^{c}_{\rho}(\eta)=\frac{G_{\rho}(\eta)}{(1-\eta)^{2h}}=\eta^{\rho}(1+a_1(\rho,c,h)\eta+a_2(\rho,c,h)\eta^2+\dots).
\end{equation}
To lighten the notation we did not show the dependence of $G_{\rho}$ from $c$ (alias $h$).
The coefficients in the series  expansion in~Eq.\eqref{conf} can be calculated  from Zamolodchikov recursive formula~\cite{Zam_rec},
see Appendix~\ref{appzam},
and provide a further verification of the  solution of Eq.~\eqref{ode}.
 Notice that, compared to~\cite{Zam_rec}, we have omitted the pre-factor $\eta^{-2h}$ in the definition of the conformal block in Eq.~\eqref{conf}.

When the indicial equation have roots $\rho_1$ and $\rho_2$, such that
$\rho_1-\rho_2=\nu$ is a positive integer but no solution is logarithmic the Frobenius series may not necessarily coincide with the conformal blocks.
The two power series can indeed mix at order $\nu$.

Since the conformal blocks are directly the contribution in Eq.~\eqref{corr_13} of the OPE channels,
we will propose an identification of the universal ratio $R$ in Eq.~\eqref{ratio_r} in terms of them. The identification is based on the following observations.
\begin{itemize}
\item[\textit{Obs. 1:}] The three linearly independent  connectivities $P_{(12)(34)}$, $P_{(14)(23)}$ and $P_{(1234)}$ are
in the scaling limit proportional to the four-point function of $\phi_{1,3}$. Therefore, once the common prefactor in Eq.~\eqref{corr_13} has been factored out,
they are linear combinations of the three conformal blocks.
\item[\textit{Obs. 2:}] The field $\phi_{1,2k+1}$ when inserted at the boundary point $x$ anchors $k$ FK-clusters~\cite{SD, Cardy_inc}.
In particular, if we consider the limit $x_1\rightarrow x_2$ in $P_{(14)(23)}$ we generate configurations where two distinct FK clusters meet
at $x_2$ and are separated by a dual FK cluster (see the previous subsection and Fig.~\ref{fig_dual}b). Such a cluster configuration is associated to the insertion at
$x_2$ of the field $\phi_{1,5}$ and therefore the leading singularity for $\eta\rightarrow 0$ of $P_{(14)(23)}$ has to be the the same as the
one of the conformal block $F^c_{3h+1}$. However $3h+1>h>0$ and we conclude then that no other conformal blocks can enter $P_{(14)(23)}$
except the one of $\phi_{1,5}$. This  observation identifies (apart from an overall constant)
the numerator in Eq.~\eqref{ratio_r}
as $(1-\eta)^{-2h}G_{3h+1}$ through Eq.~\eqref{conf}.
\item[\textit{Obs. 3:}] Consider the OPE of two bcc operators $\phi_{\alpha\beta}$ as they appear in Eq.~\eqref{bcc_conn}, it has the structure
\begin{equation}
\label{OPE}
\phi_{\alpha\beta}\cdot\phi_{\beta\alpha}=\mathbf{1}+\mathbf{X}+\dots,
\end{equation} 
where $\mathbf{1}$ is the identity field and $\mathbf{X}$ denotes a field with scaling dimension larger than zero that is compatible with the boundary conditions. 
Certainly such a field cannot be
$\phi_{\alpha\beta}$, since this would imply a discontinuity of the boundary conditions that is not allowed by the OPE in Eq.~\eqref{OPE},
see also \cite{DVPotts} for analogous arguments for kink fields in the bulk. Therefore we are led to the
conclusion that the conformal blocks that enter the denominator in Eq.~\eqref{ratio_r} can only be  $F_{0}^c$ (i.e. the identity conformal block) and
$F_{3h+1}^c$ (i.e. the $\phi_{1,5}$ conformal block).
\item[\textit{Obs. 4:}] The denominator in Eq.~\eqref{ratio_r} is obviously symmetric under the exchange $x_1\rightarrow x_3$, Such a symmetry corresponds
to the transformation $\eta\rightarrow(1-\eta)$. Imagine now, according to the previous observation, to have expressed the denominator in
Eq.~\eqref{ratio_r} as a linear combination ($\alpha,\beta\in\mathbb R$)
\begin{equation}
\label{symS}
 \alpha F_{0}^c(\eta)+\beta F^{c}_{3h+1}(\eta)\equiv\frac{S(\eta)}{(1-\eta)^{2h}},
\end{equation}
then it easy to verify, substituting into Eq.~\eqref{corr_13} that the symmetry under the exchange $x_1\rightarrow x_3$ requires the function
$S$ to satisfy
the functional equation $S(\eta)=S(1-\eta)$. Obviously $S$ is a solution of Eq.~\eqref{ode} and from the fusion matrix given in~\cite{DF} (see Eq. (5.11) there)
it can be moreover verified that it exists only a linear independent function $S$, constructed from the conformal blocks of the identity and $\phi_{1,5}$
that satisfies such a property (the other can be chosen a linear combination of the conformal blocks of $\phi_{1,3}$ and $\phi_{1,5}$).  
\end{itemize}

In summary the universal ratio $R$ in Eq.\eqref{ratio_r} will be 
\begin{equation}
\label{ratiofin}
R(\eta)=A_Q\frac{G_{3h+1}(\eta)}{S(\eta)},
\end{equation}
where $A_Q$ imposes $R(1)=1$  that should be clear from the geometrical interpretation in Fig.~\ref{fig_conn}.
\section{Analytic expressions for the connectivities}
\label{sec3}
\begin{figure}[t]
\centering
\includegraphics[width=0.65\textwidth]{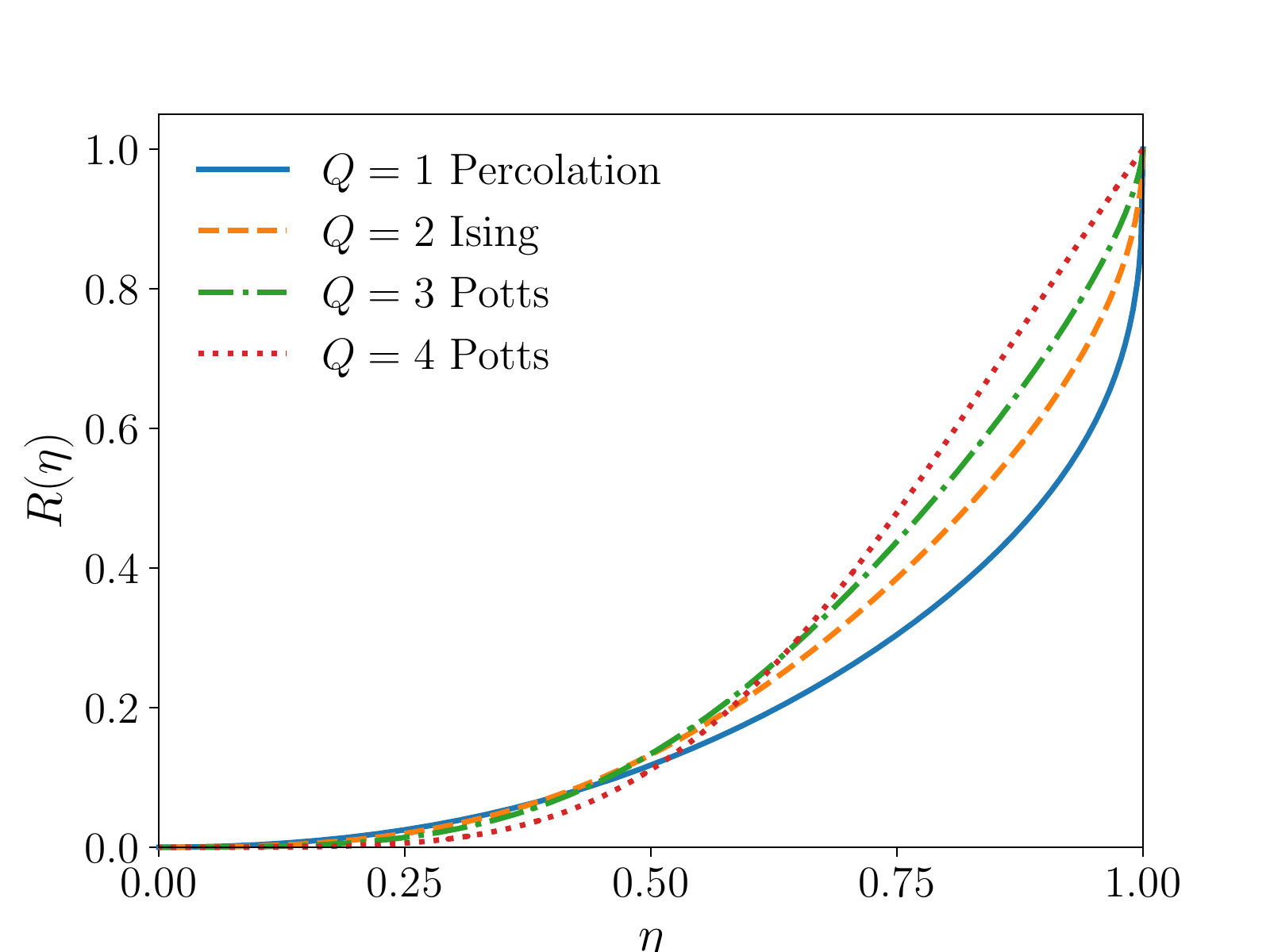}
\caption{The universal ratio $R$ in Eq.~\eqref{ratio_r}  in the critical $Q$-color Potts model as a function of the anharmonic ratio $\eta$.}
\label{figrs}
\end{figure}
We are ready to present explicit results for the ratio $R$ in Eq.~\eqref{ratio_r} for
any integer $1\leq Q\leq 4$. When needed to distinguish among different $Q$'s in Eq.~\eqref{ratio_r}
we will introduce an extra index $Q$
(see for instance Eq.~\eqref{ratioperc} or Eq.~\eqref{ratio_ising}). The functions $R_Q$, as derived in the
following, are shown in Figure~\ref{figrs}.
Finally, for more technical details,
we invite to read the three final Appendices. 
\newline
\textit{Q=1; Percolation---} We have (cf. Eq.~\eqref{cc}) $p=2$ and $h_{1,3}=h=1/3$, $h_{1,5}=3h+1=2$. The solutions $G_{1/3}(\eta)$ and $G_2(\eta)$ are free
of logarithms and can be easily determined using the method described in Appendix~\ref{app}.  It turns out
\begin{align}
\label{13Q1}
G_{1/3}(\eta)=\eta^{1/3}\left(1-\frac{1}{2}\eta-\frac{2}{7}\eta^2-\frac{1}{7}\eta^3-\frac{58}{637}\eta^4-\frac{83}{1274}\eta^5+o(\eta^5)\right),\\
\label{2Q1}
G_2(\eta)=\eta^2\left(1+\frac{1}{3}\eta+\frac{37}{198}\eta^2+\frac{112}{891}\eta^3+\frac{469}{5049}\eta^4+\frac{3304}{45441}\eta^5+o(\eta^5)\right);
\end{align}
and we actually generated $O(10^5)$ terms in all the power series.
The conformal blocks $F^{0}_{1/3}(\eta)$ and $F^{0}_{2}(\eta)$ are obtained through Eq.~\eqref{conf} as it can be verified from Appendix~\ref{appzam}. The conformal block of the identity field is  singular at $c=0$~\cite{GL}
and needs to be regularized. In particular, Eq.~\eqref{symS} does not hold \textit{verbatim} at $c=0$; although it might be still true
in the limit $Q\rightarrow 1$. Here, we  proceed
pragmatically finding the Frobenius solution with $\rho=0$ of Eq.~\eqref{ode}, that is symmetric under the exchange
$\eta\rightarrow (1-\eta)$; such a solution is logarithmic. Using the method described in Appendix~\ref{app} and
in particular the normalization $b_0(\sigma)=\sigma$ we obtain
\begin{equation}
\label{log}
\tilde{G}_0(\eta)=-\frac{8}{45}\log(\eta) G_{2}(\eta)+\left(1-\frac{2}{3}\eta+\frac{119}{225}\eta^2+\frac{152}{2025}\eta^3+\frac{18947}{735075}\eta^4+\frac{27058}{2205225}\eta^5+o(\eta^5)\right).
\end{equation}
Remarkably, the three power series in Eqs.~(\ref{13Q1}-\ref{log}) can be  expressed through combinations of
hypergeometric functions $_{3}F_{2}$; see Eqs.~(\ref{g0_hyper}-\ref{g1_hyper})
or Eqs.~(\ref{g0series}-\ref{g2series}). In particular we can construct directly a
symmetric solution of Eq.~\eqref{ode}, whose series expansion near $\eta=0$ coincides  with Eq.~\eqref{log}, thus proving that
$\tilde{G}_{0}(\eta)=\tilde{G}_0(1-\eta)$.

Notice also that the second linearly independent solution symmetric under the transformation $\eta\rightarrow (1-\eta)$  can be
chosen $G_{1/3}(\eta)+G_{1/3}(1-\eta)$. However such a function cannot enter into $S(\eta)$ since it
contains a subleading singularity $\eta^{1/3}$. In conclusion, we conjecture that at $Q=1$, $S(\eta)=\tilde{G}_0(\eta)$,
given in Eq.~\eqref{log}.

It is also important to mention that the coefficient of the logarithm in
Eq.~\eqref{log} is related to the Gurarie-Ludwig~\cite{GL} indecomposability parameter $b$.
It was indeed argued that CFTs at $c=0$ could be characterized by a universal number $b$ appearing
in the regularized OPE of a chiral field with itself, see Eq.~\eqref{OPE1}.  In particular if $h$ is the (chiral) scaling dimension of such a field,
the leading small $\eta$ behavior of the function $\tilde{G}_{0}$
(notice the definition of the prefactor in Eq.~\eqref{corr_13}) has to be~\cite{GL} 
\begin{equation}
\tilde{G}_0(\eta)=1-2h\eta+\frac{h^2}{b}\eta^2\log(\eta)+...,
\end{equation}
Comparing with Eq.~\eqref{log} for $h=1/3$  we obtain  $b=-5/8$. This is indeed the
same value of the indecomposability parameter that was argued to describe critical boundary percolation~\cite{DJS}. However no logarithmic observable in critical two-dimensional percolation was fully calculated so-far.
Summarizing we have (cf. Eq.~\eqref{ratio_r})
\begin{equation}
\label{ratioperc}
R_{Q=1}=A_1\frac{G_2(\eta)}{\tilde{G}_0(\eta)},
\end{equation}
and $A_1$  ensures $R_{Q=1}(\eta=1)=1$; see Appendix ~\ref{hyper} for an explicit expression for
Eq.~\eqref{ratioperc}, including the constant $A_1$. \newline
\noindent
\textit{Q=2; Ising model---} Here $p=3$ and $h_{1,3}=h=1/2$, $h_{1,5}=3h+1=5/2$. This case was solved in~\cite{GV17}, however we report it for completeness.
The power series solution $G_0(\eta)$ and $G_{5/2}$ are
\begin{align}
\label{ising0}
&G_0(\eta)=G_0(1-\eta)=1-\eta+\eta^2,\\
\label{ising52}
&G_{5/2}(\eta)=\eta^{5/2}\left(1+\frac{1}{4}\eta+\frac{49}{384}\eta^2+\frac{125}{1536}\eta^3+\frac{37025}{638976}\eta^4+\frac{37547}{851968}\eta^5+o(\eta^5)\right).
\end{align}
and are related to the conformal blocks $F^{1/2}_0(\eta)$ and $F^{1/2}_{5/2}(\eta)$  through Eq.~\eqref{conf}.
Since the function $G_0(\eta)$ is symmetric,  it follows from Eq.~\eqref{symS} $G_0(\eta)=S(\eta)$; i.e. at $c=1/2$,
the coefficient $\beta$ in the linear combination in Eq.~\eqref{symS} is zero which simplified the discussion in~\cite{GV17}.
It can also be proven that  the power series for $G_{5/2}$ can be re-summed as follows
\begin{equation}
G_{5/2}(\eta)=G_0(\eta)\int_0^{\eta}d\eta'g(\eta'),
\end{equation}
being~\cite{key, GV17}
\begin{equation}
\label{f}
g(\eta)=\frac{16}{21 \pi}\frac{(2-\eta) (1+\eta) 
 (-1+2 \eta) E(\eta)+(2+\eta (-4+\eta+\eta^2)) K(\eta)}{\sqrt{(1-\eta) \eta} (1+(-1+\eta) \eta)^2}.
\end{equation}
In Eq.~\eqref{f} above $E(\eta)$ and $K(\eta)$ are the 
complete elliptic integrals of first and second kind
(with \texttt{Mathematica} convention for the modulus). In conclusion~\cite{GV17} the ratio $R$ at $Q=2$ is given by
\begin{equation}
\label{ratio_ising}
R_{Q=2}=A_2\frac{G_{5/2}(\eta)}{G_{0}(\eta)}=A_2\int_0^{\eta}d\eta'g(\eta');
\end{equation}
where $A_2$ ensures $R_{Q=2}(1)=1$. The logarithmic behavior emerges in Eq.~\eqref{ratio_ising} in the limit
$\eta\rightarrow 1$ and algebraically
is understood by the collision of the primary field $\phi_{1,5}$ with the null vector at level $2$ of $\phi_{1,3}$~\cite{GanVa, SV}. \newline
\noindent
\textit{Q=3; Three-color Potts model---}The model corresponds to $p=5$ and $h_{1,3}=h=2/3$, $h_{1,5}=3h+1=3$. There are no logarithmic solutions; using the method of Appendix~\ref{app} we obtain the power series
\begin{align}
\label{symm3}
&G_0(\eta)=G_0(1-\eta)=1-\frac{4}{3}\eta+\frac{4}{3}\eta^2,\\
&G_{2/3}(\eta)=\eta^{2/3}(1-\eta+\frac{3}{4}\eta^2),\\
&G_{3}(\eta)=\frac{81}{52}\left[G_0(\eta)-\frac{4}{3}G_{2/3}(1-\eta)\right].
\end{align}
The conformal blocks  $F_{2/3}^{4/5}(\eta)$ and $F_{3}^{4/5}(\eta)$ are obtained from Eq.~\eqref{conf}. Using the recursive formula in Appendix~\ref{appzam} we can also verify
directly that $G_0(\eta)(1-\eta)^{-4/3}=F_0^{4/5}(\eta)+\frac{26}{81}F_{3}^{4/5}(\eta)$, thus proving that $G_0(\eta)$ in Eq.~\eqref{symm3} is
actually $S(\eta)$; cf. Eq.~\eqref{symS}. We therefore
have at $Q=3$ 
\begin{equation}
\label{r3}
R_{Q=3}=A_3\frac{G_3(\eta)}{G_0(\eta)}=1-\frac{(1-\eta)^{2/3} \left(1-\frac{2 \eta}{3}+\eta^2\right)}{1-\frac{4 \eta}{3}+\frac{4 \eta^2}{3}}.
\end{equation}
\newline
\noindent
\textit{Q=4; Four-color Potts model---}Finally we consider the four-color Potts model for which $p\rightarrow\infty$ and $h_{1,3}=h=1$, $h_{1,5}=3h+1=4$.
All the Frobenius power series reduce to polynomials; Appendix \ref{app} produces the following basis of solutions: $G_0(\eta)=1$, $G_1(\eta)=\eta-\frac{3}{2}\eta^2+\eta^3$, $G_4(\eta)=\eta^4$.
However, as we remarked in Sec.~\ref{sec3} only the conformal block $F^{1}_4$ is obtained from $G_4$ using Eq.~\eqref{conf}. The conformal blocks of the identity and
the field $\phi_{1,3}$ are derived from the recursive formula in Appendix~\ref{appzam}. To calculate the identity conformal block  from the recursive formula, the limit $c\rightarrow 1$ must be taken after sending the internal conformal dimension of the block to zero. This is the correct way of avoiding contributions from spurious states with zero norm in the corresponding Virasoro algebra irreducible representation~\footnote{We thank the  referee for showing this to us.}. The results are
\begin{align}
\label{idq4}
 &F_0^{1}(\eta)=\frac{G_0(\eta)-2G_1(\eta)+\frac{1}{3}G_4(\eta)}{(1-\eta)^2},\\
 &F_{1}^{1}(\eta)=\frac{G_1(\eta)-\frac{1}{4}G_{4}(\eta)}{(1-\eta)^2}.
\end{align}
According to Eq.~\eqref{symS} there exists only a linear combination of $F_{0}^1$ and $F_{4}^1$ that leads to a function $S$ symmetric under the
transformation $\eta\rightarrow(1-\eta)$. We indeed find
$F_0^1+\frac{2}{3}F_{4}^1=(1-\eta)^{-2}(1-\eta+\eta^2)^2$ and therefore $S(\eta)=(1-\eta+\eta^2)^2$. Consistently with the discussion below Eq.~\eqref{symS} the
other linear combination leading to a symmetric function can be chosen $F_1^{1}-\frac{1}{4}F_{4}^1$. Summarizing at $Q=4$
\begin{equation}
\label{ratio4}
R_{Q=4}=A_4\left(\frac{\eta^2}{1-\eta+\eta^2}\right)^2=\left(\frac{\eta^2}{1-\eta+\eta^2}\right)^2.
\end{equation} 

\section{Numerical Experiments}
\label{sec4}
\begin{figure}[t]
\centering
\includegraphics[width=0.5\textwidth]{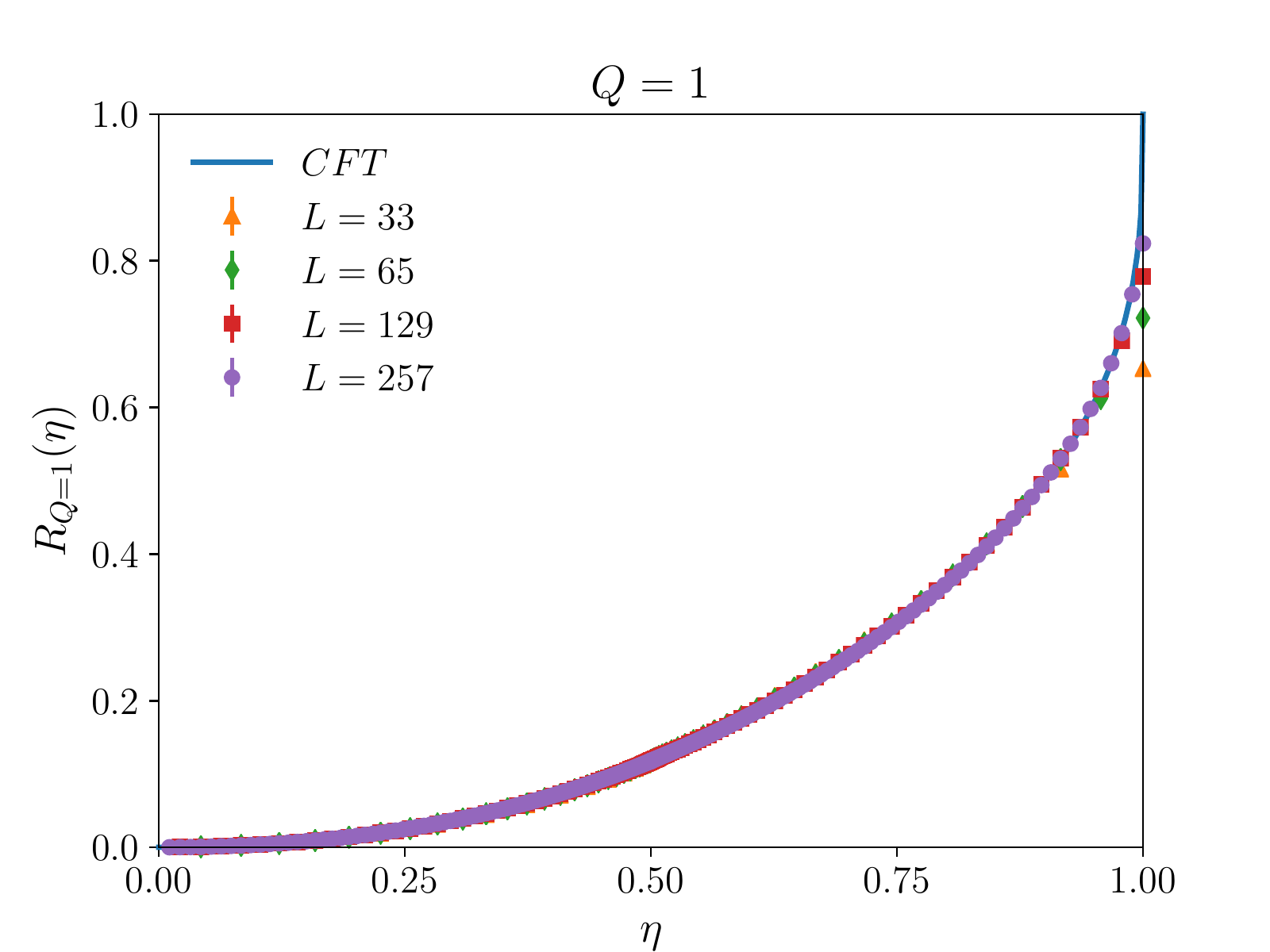}~
\includegraphics[width=0.5\textwidth]{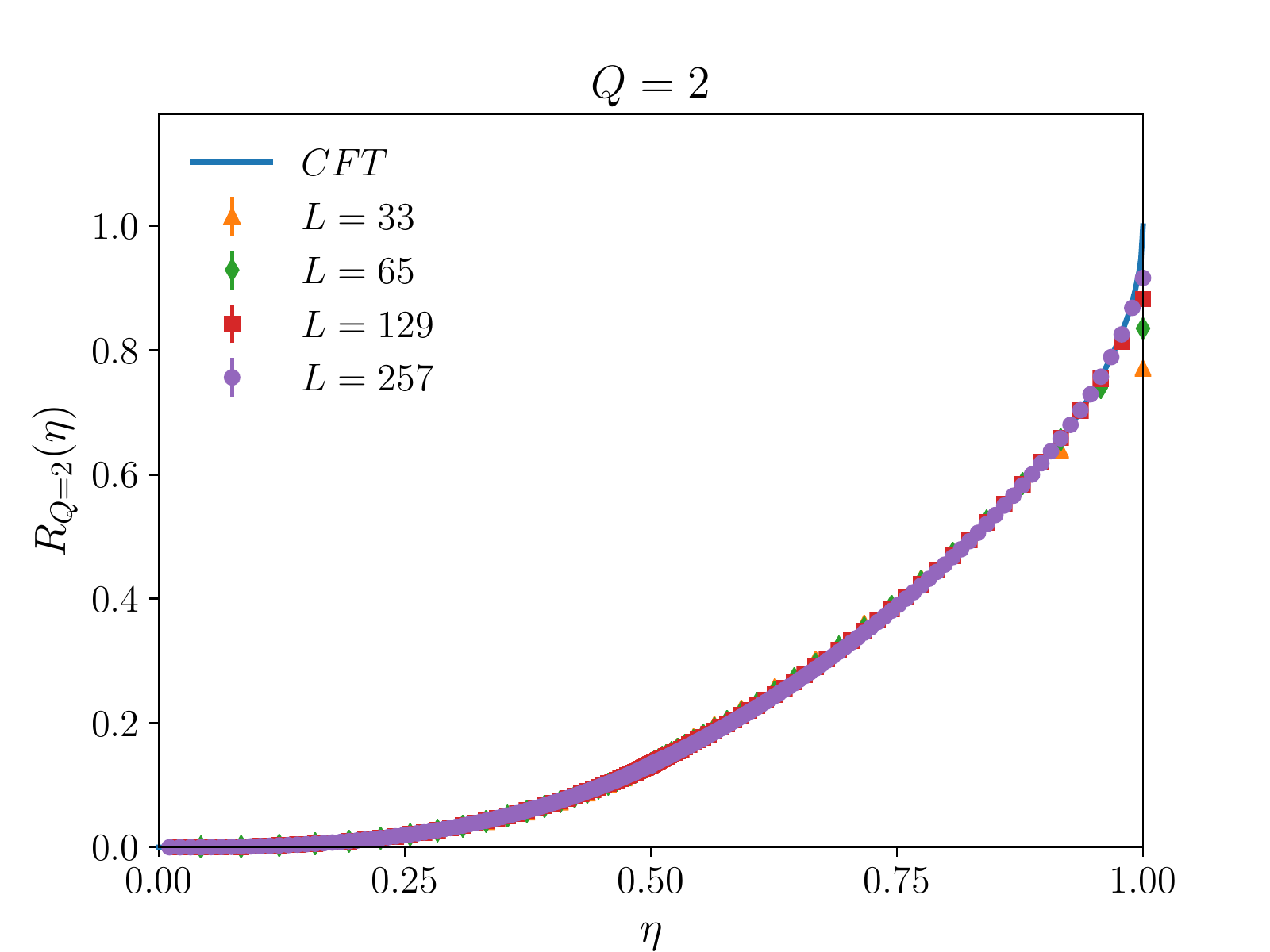}
\includegraphics[width=0.5\textwidth]{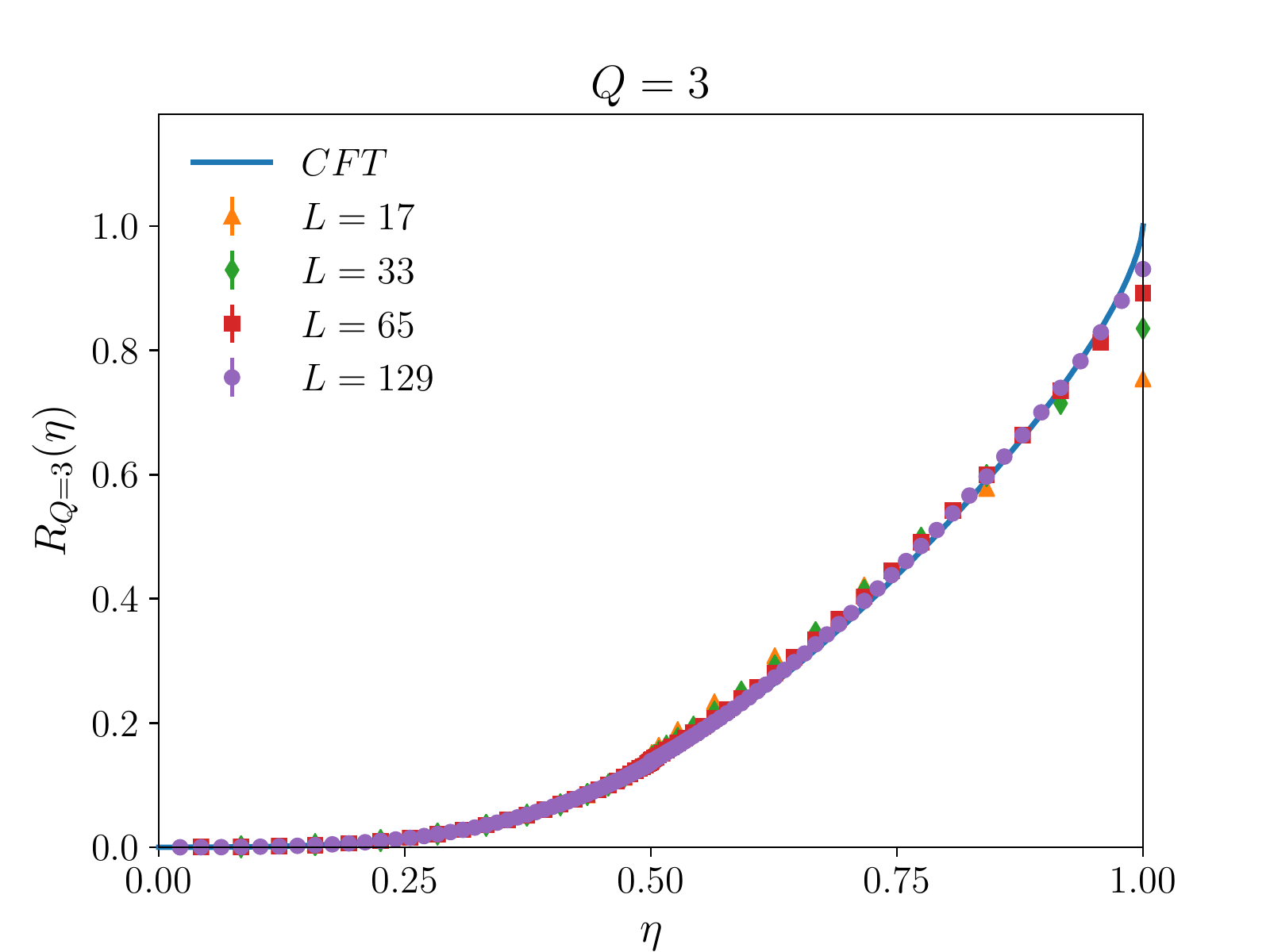}~
\includegraphics[width=0.5\textwidth]{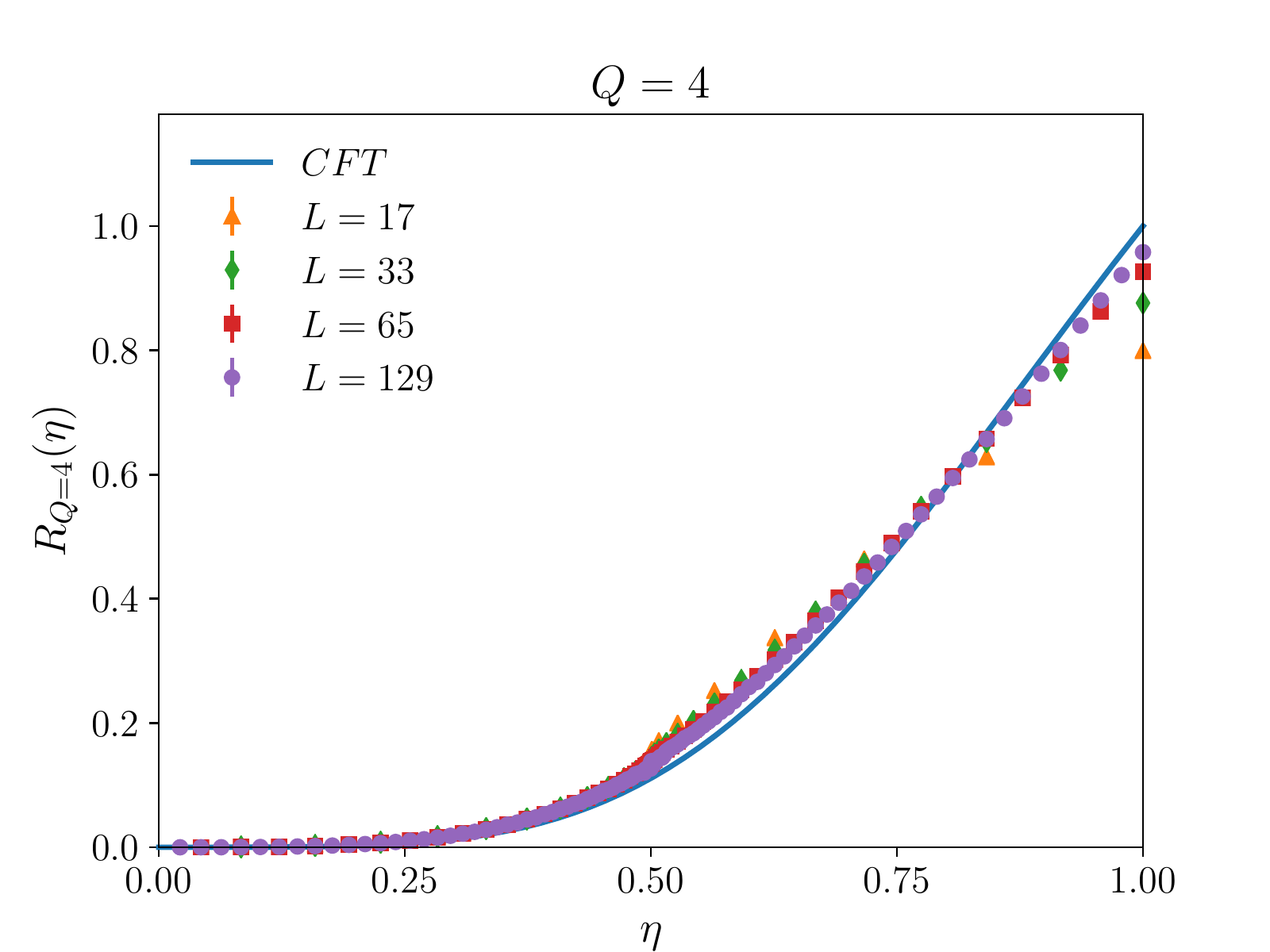}
\caption{The plots refer to percolation (top left), Ising (top right), three color (bottom left) and four color Potts (bottom right) model respectively and show the results of Monte Carlo simulations for the ratio $R^{MC}_Q$ for sizes $L=33$, $65$, $129$, $257$ for $Q=1$, $2$ and $L=17$, $33$, $65$, $129$
for $Q=3$, $4$ together with the theoretical CFT predictions $R^{CFT}_Q$.
}
\label{fig_func}
\end{figure}
The conjectures presented in Sec.~\ref{sec3} have been tested against
 extensive Monte Carlo simulations. We give some details about the numerical experiments.
Simulation have been carried on triangular lattices on triangles of side $L$ where
$L=33$, $65$, $129$, $257$. The value of the reduced inverse temperature has been set to the
exactly known~\cite{Wu} critical value in the thermodynamic limit 
\begin{equation}
\label{cp}
e^{2 J_c} = 2 \cos \left[\frac{2}{3} \arccos\left(\frac{\sqrt{Q}}{2}\right)\right].
\end{equation}
As done in~\cite{GV17}, we map the four points $z_1=0, z_2=\eta,~z_3=1$ and $z_4=\infty$ on the boundary
of an equilateral triangle by a Schwartz-Christoffel transformation 
\begin{equation}
\label{sc}
 w(z)=\frac{6z\Gamma\left(\frac{5}{6}\right){}_2 F_1(1/2,2/3; 3/2, 9z^2)}{\sqrt{\pi}\Gamma\left(\frac{1}{3}\right)}.
\end{equation}
In particular, the points $z_1=0$, $z_3=1$ are mapped through Eq.~\eqref{sc} into the
midpoints of an equilateral triangle with length-side two and vertices at $w(-1/3)=-1$, $w(1/3)=1$ and $w(\infty)=-i\sqrt{3}$.  The image of the point
$z_2$ moves therefore along the triangle between $w(z_1)=0$ and $w(z_2)=e^{-i\pi/3}$.  Symmetries of the triangle are also taken into account in
order to enhance the statistics. The algorithm employed is the Swendsen-Wang cluster
algorithm~\cite{swendsen_1987} giving direct access to the FK clusters.
The random number generator is given 
in~\cite{matsumoto_1998} and the number of samples collected is up to $10^{10}$ for
the largest sizes considered. As the size is increased all crossing events become
rarer and this happens in a more severe way for higher values of the parameter $Q$.
This can obviously be traced back to the leading scaling dimension $h_{1,3}$ setting the dimensions of the numerator and denominator
in Eq.~\eqref{ratio_r}. Its value  gets bigger as $Q$ is increased. This has limited the maximal size of the triangular lattice for
$Q=3$, $4$ to $L=129$. 

\begin{figure}[t]
\centering
\includegraphics[width=0.5\textwidth]{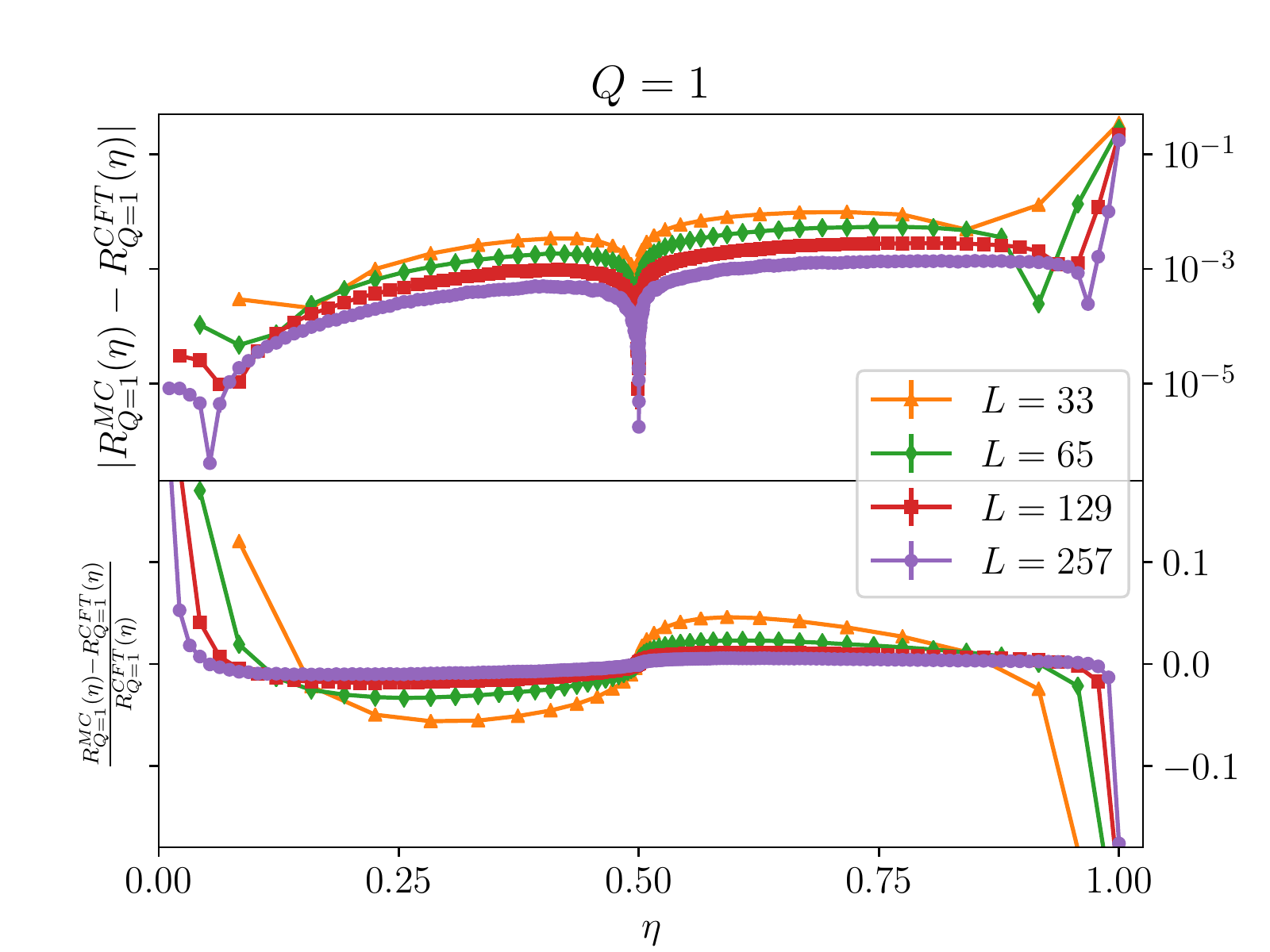}~
\includegraphics[width=0.5\textwidth]{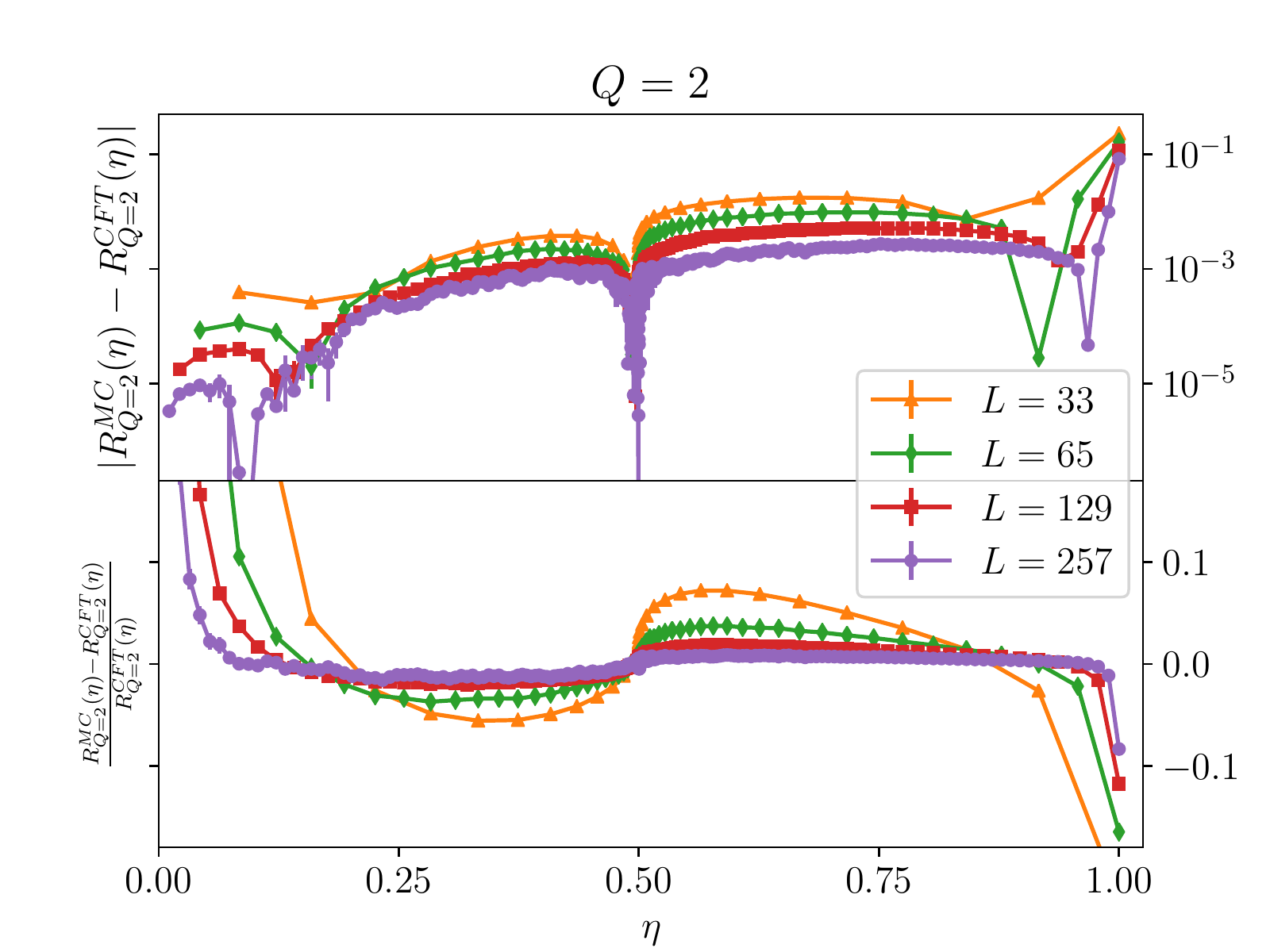}
\includegraphics[width=0.5\textwidth]{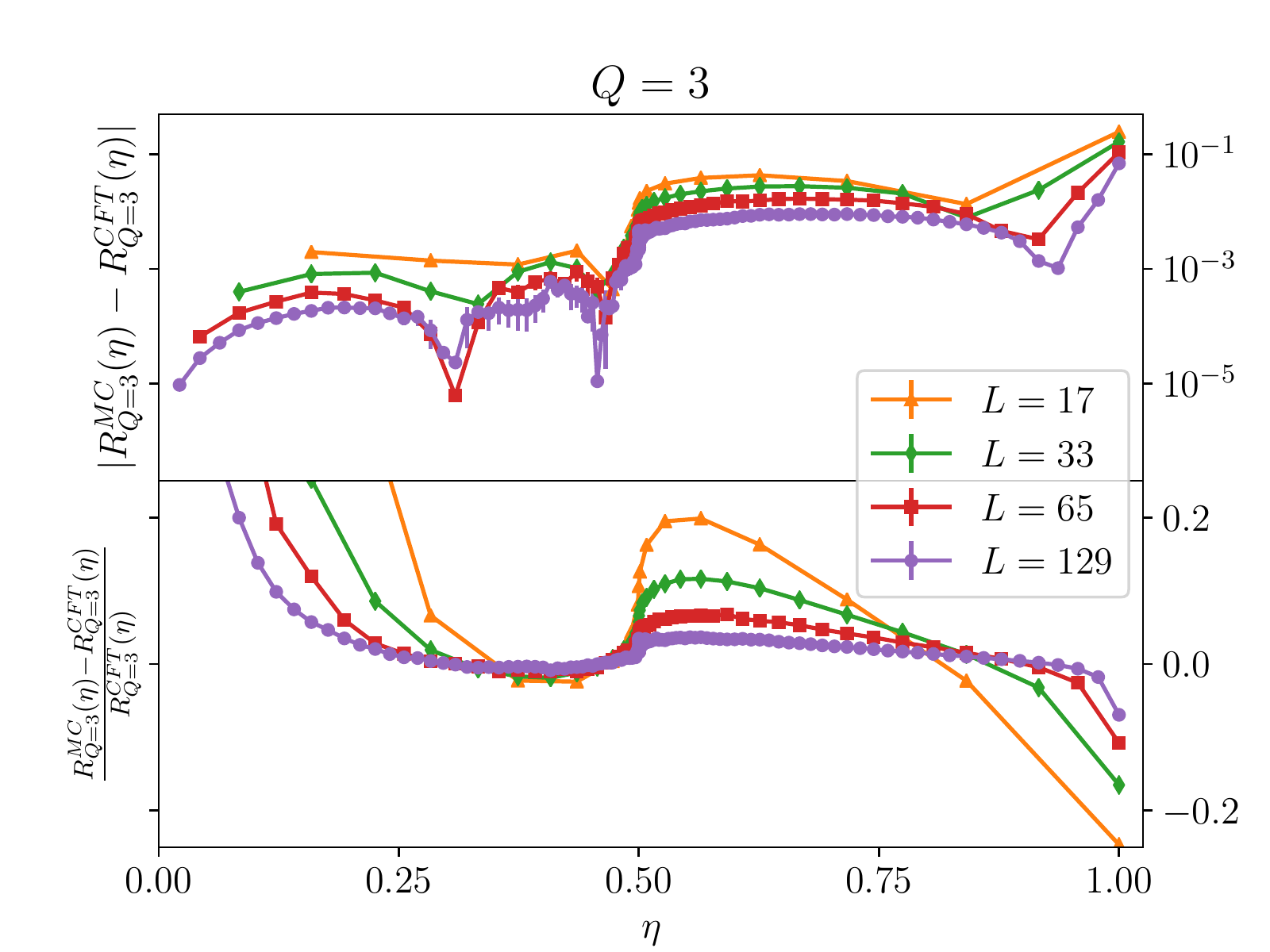}~
\includegraphics[width=0.5\textwidth]{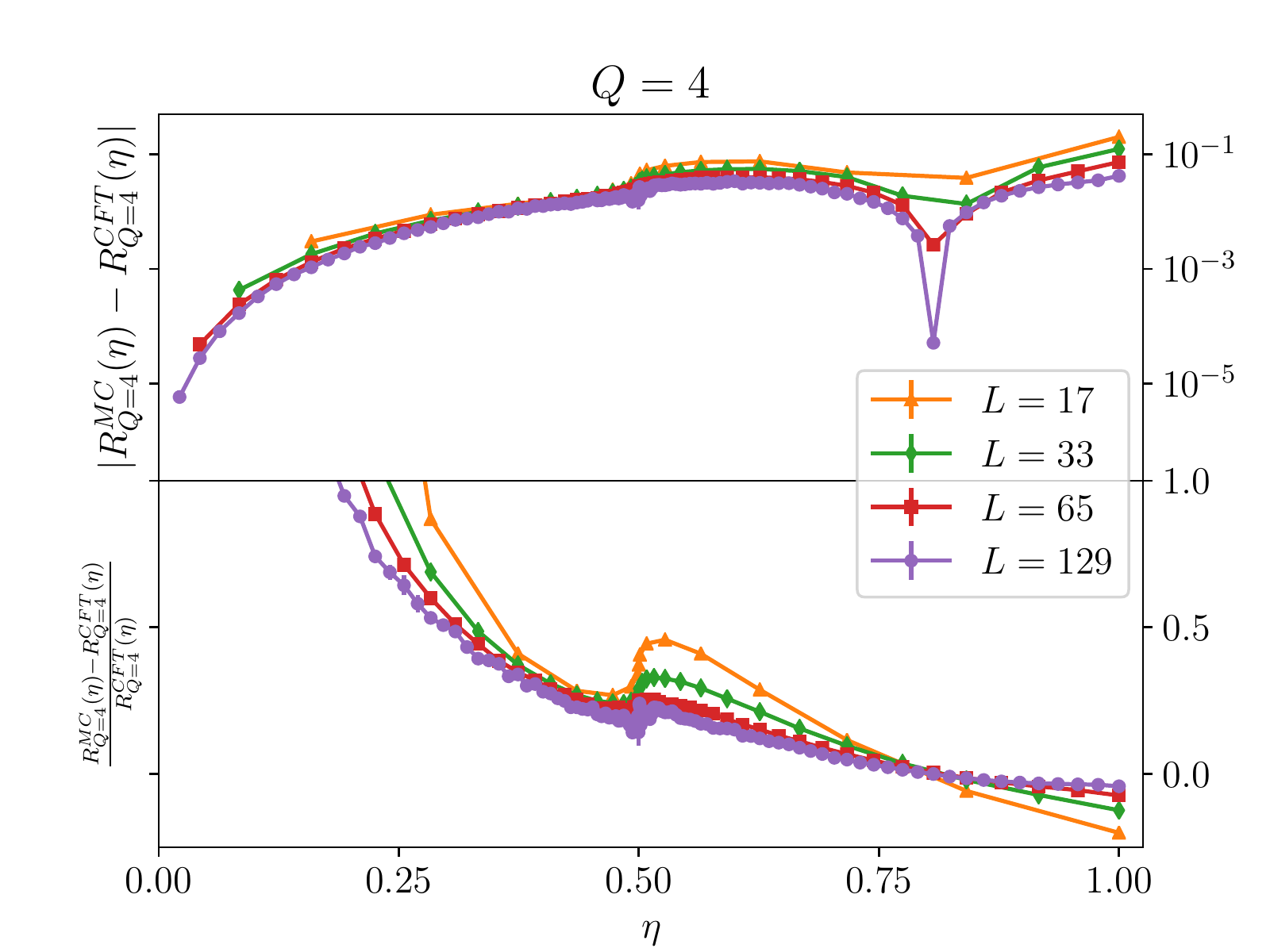}
\caption{ 
The upper part of each plot  shows the absolute value of the deviations from the CFT predictions discussed in Sec.~\ref{sec3}
(notice the logarithmic scale) while the lower
part displays  the signed relative deviations.
}
\label{fig_diff}
\end{figure}

\begin{figure}[t]
\centering
\includegraphics[width=0.5\textwidth]{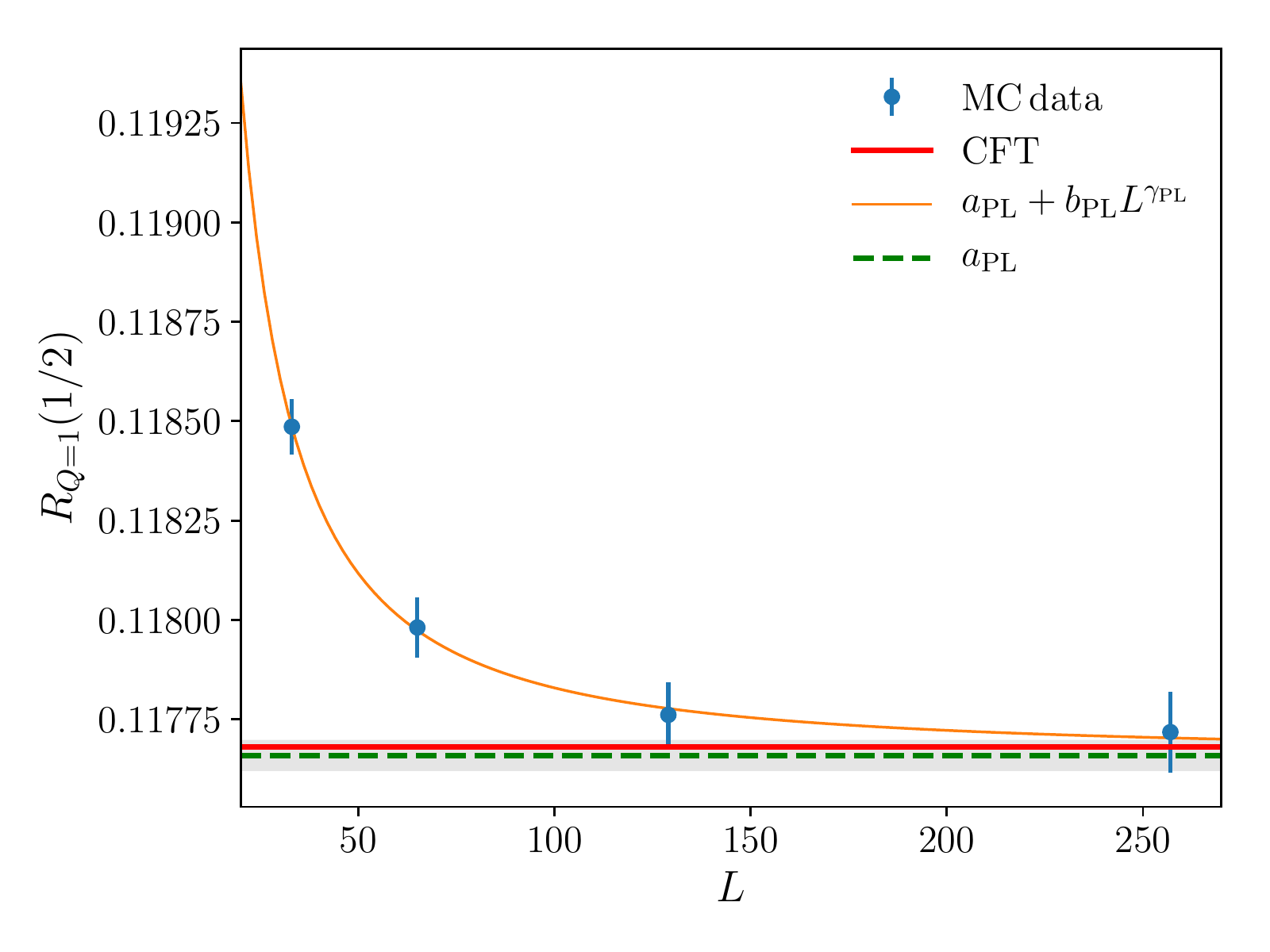}~
\includegraphics[width=0.5\textwidth]{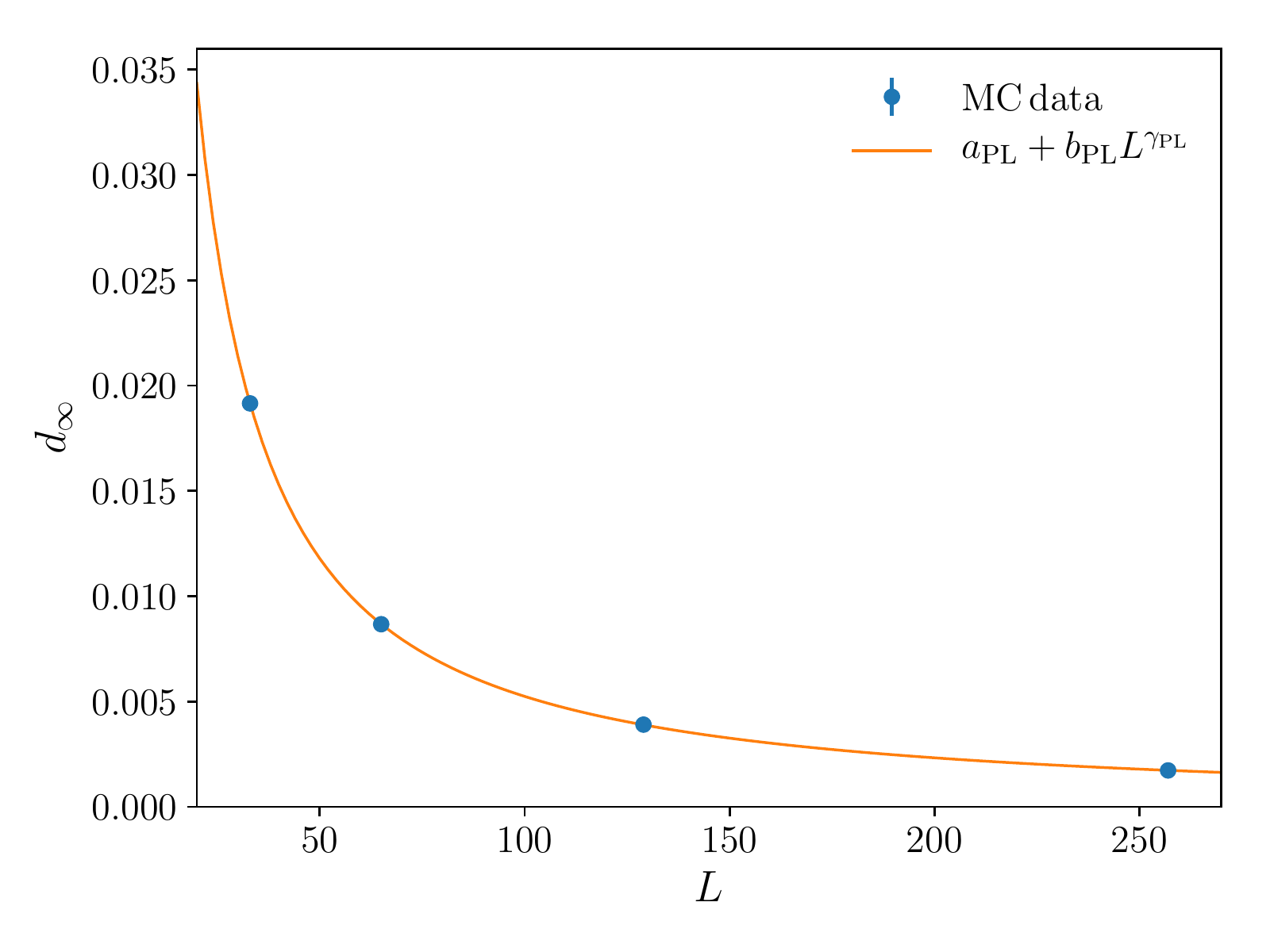}
\caption{(Left panel) Value of $R^{MC}_{Q=1}(1/2)$ as a function of the system size $L$.
The yellow curve is a fit with a power law model. The dotted green line is the fitted value
for $a_\mathrm{PL}$, with the shaded area representing the confidence interval. The red line is the CFT prediction.
(Right panel) Distance $d_\infty$ (see Eq.~\eqref{dinfty}) between the Monte Carlo measured $R^{MC}_{Q=1}$ and the
CFT prediction as a function of $L$. The data can be fitted with a power law function $a_\mathrm{PL}+b_\mathrm{PL}L^{\gamma_\mathrm{PL}}$ with $a_\mathrm{PL}=(-1\pm3)\cdot 10^{-5}$.
}
\label{fig_q1}
\end{figure}

 The data from the simulations are presented in Fig. \ref{fig_func}. We first note that the simulation data fall on the predicted curves with good and increasing accuracy as the system size increases. In Fig. \ref{fig_diff} we show the deviations from the CFT conjectures for the different sizes considered. The main plots show the absolute values of deviations (notice the logarithmic scale) while the insets refer to the signed relative deviations. As it can be observed, a non-trivial behavior emerges. Some features are easily explained: the deviations near the endpoints, especially apparent in the relative deviations plot, are obviously
due to short-distance lattice effects, indeed their extent is confined to smaller regions as the size is increased. Extrapolation of the ratio $R$ (Eq.~\eqref{ratio_r}) in thermodynamic limit would require a theory of finite size corrections which is largely unknown and beyond the purposes of this paper. 
In order to give a quantitative assessment of the convergence to our theoretical predictions we employ
the $L^{\infty}$ norm in the space of functions.  The distance between the Monte Carlo data $R^{MC}_Q(\eta)$
and the CFT prediction $R^{CFT}_{Q}(\eta)$ is then defined by
\begin{equation}
\label{dinfty}
 d_\infty=\int_0^1 \mathrm{d}\eta |R^{MC}_{Q}(\eta)-R^{CFT}_{Q}(\eta)|.
\end{equation}
The data obtained will be extrapolated to the thermodynamic limit by fitting the $L$ dependence 
through the following Ans\"atze:   
A power law of the  form $a_\mathrm{PL}+b_\mathrm{PL} L^{\gamma_\mathrm{PL}}$ and only at $Q=4$  (see later) a logarithmic scaling
of the form $a_{\log}+b_{\log} \log(L)^{\gamma_{\log}}$.
Equipped with these tools we discuss now the different values of $Q$ considered.
Since they will display qualitatively different behaviors we will separate the $Q=1,2,3$ cases
from  $Q=4$.
\\
\noindent
\textit{$Q=1,2,3$ cases---} 
Percolation ($Q=1$) exhibits a remarkable convergence to
the conjectured formula (Eq.~\eqref{ratioperc}). The numerical data approach the thermodynamic
limit in a smooth way (see Fig.~\ref{fig_func}). Due to this regularity 
we can also try to inspect the pointwise convergence. As a case study, we concentrate on the point $\eta=1/2$. If we extrapolate to the thermodynamic limit the numerically obtained values 
by fitting them with a power law model $R^{MC}_{Q=1}(1/2)=a_\mathrm{PL}+b_\mathrm{PL} L^{\gamma_\mathrm{PL}}$ we get
$a_\mathrm{PL}=0.11766(4)$ fully consistent with the conjectured expression that yields
$R_{Q=1}(1/2)=0.117680185\ldots$ (see Fig.~\ref{fig_q1} left panel). This is the same kind of agreement obtained 
in the most precise tests of CFT (see for example \cite{CZ}
where a result correct to six significant digits for the coefficient of universal area
distribution of clusters in percolation is obtained).
Let us now turn to the more relevant global convergence:
As can be seen in the right panel of Fig.~\ref{fig_q1} the distance $d_\infty$ strikingly converges,
with a power law decay, to a value compatible with zero within small errors.

\begin{figure}[t]
\centering
\includegraphics[width=0.5\textwidth]{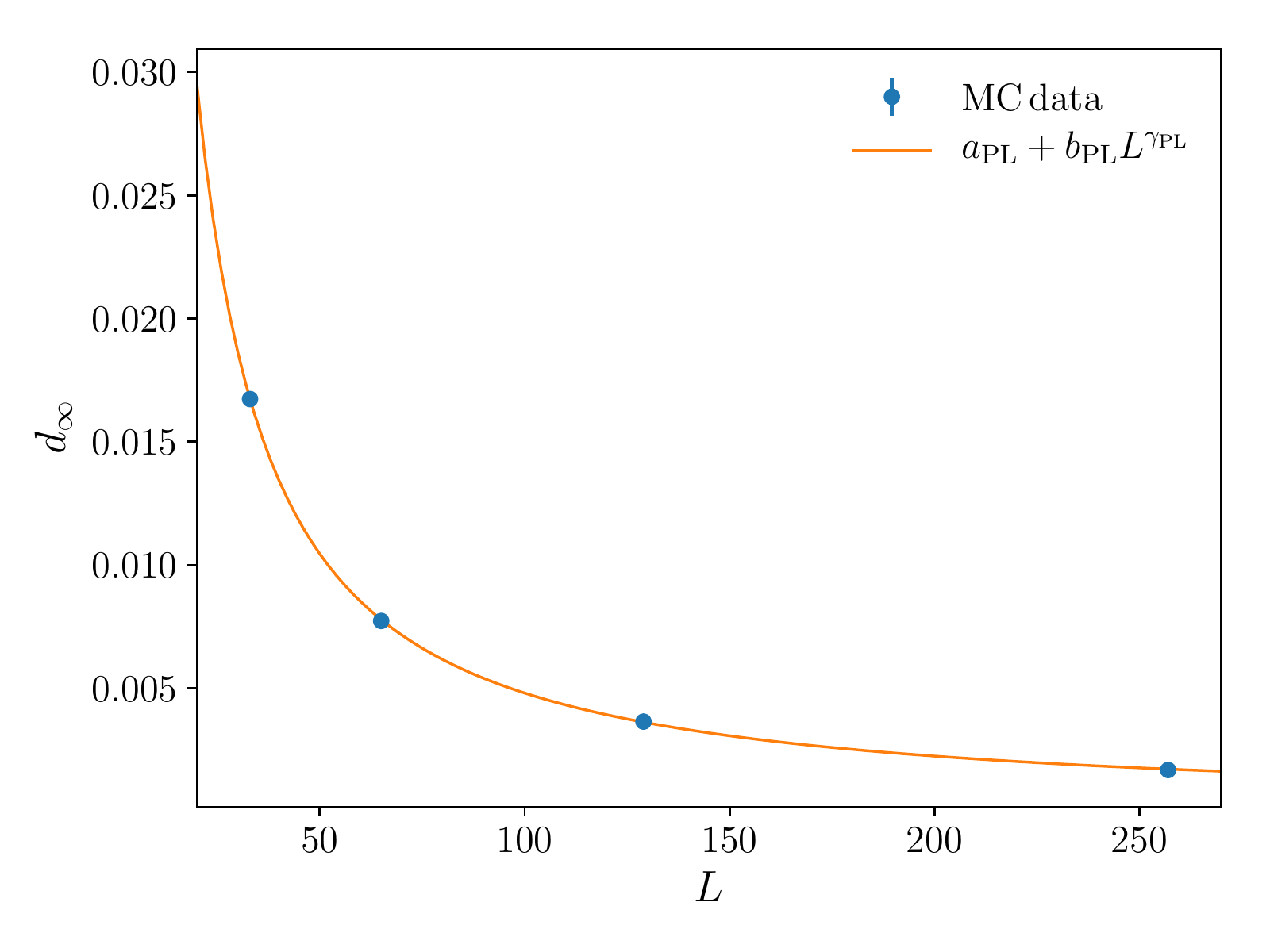}~
\includegraphics[width=0.5\textwidth]{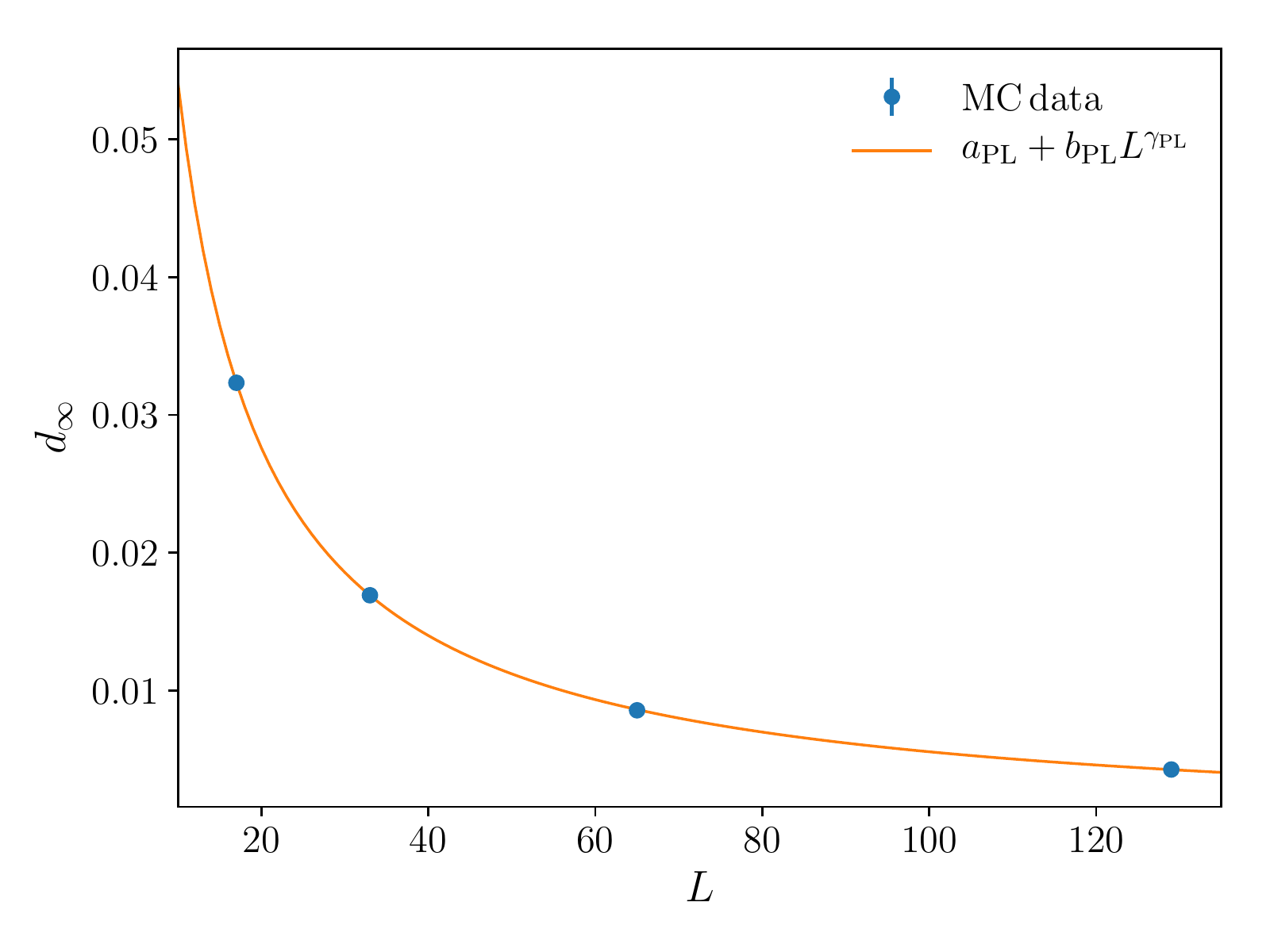}
\caption{Distance $d_\infty$ (see Eq.~\eqref{dinfty}) between the measured  $R_Q^{MC}$ and the
CFT predictions as a function of $L$ for $Q=2$ (left) and $Q=3$ (right). The data can be 
fitted with a power law function $a_\mathrm{PL}+b_\mathrm{PL}L^{\gamma_\mathrm{PL}}$
with $a_\mathrm{PL}=(1.0\pm1.4)\cdot 10^{-4}$ and $a_\mathrm{PL}=(-4.3\pm0.8)\cdot 10^{-4}$ for $Q=2$, $3$ respectively.
}
\label{fig_q2_q3}
\end{figure}

Now we turn to the cases $Q=2$ and $Q=3$. Concerning pointwise convergence, 
we observe that the curves become more complex (especially for $Q=3$), alternating 
regions that overshoot or undershoot the predicted curve. In addition these regions
move as the size is increased making more difficult a phenomenological description,
lacking the support of a theory of finite size corrections for the problem
under scrutiny. When we examine instead $L^\infty$ convergence (shown in Fig.~\ref{fig_q2_q3}) 
the value of the distance extrapolates to zero with high precision, 
supporting a convincing convegence to our  formulas in Eqs.~\eqref{ratio_ising} and \eqref{r3}. The result for the $Q=3$ Potts model also suggests that small subleading corrections to the scaling should be taken into account.
In conclusion, the Monte Carlo study leaves little room for doubts about the validity of the CFT predictions also  at $Q=2,3$.

\begin{figure}[t]
\centering
\includegraphics[width=0.5\textwidth]{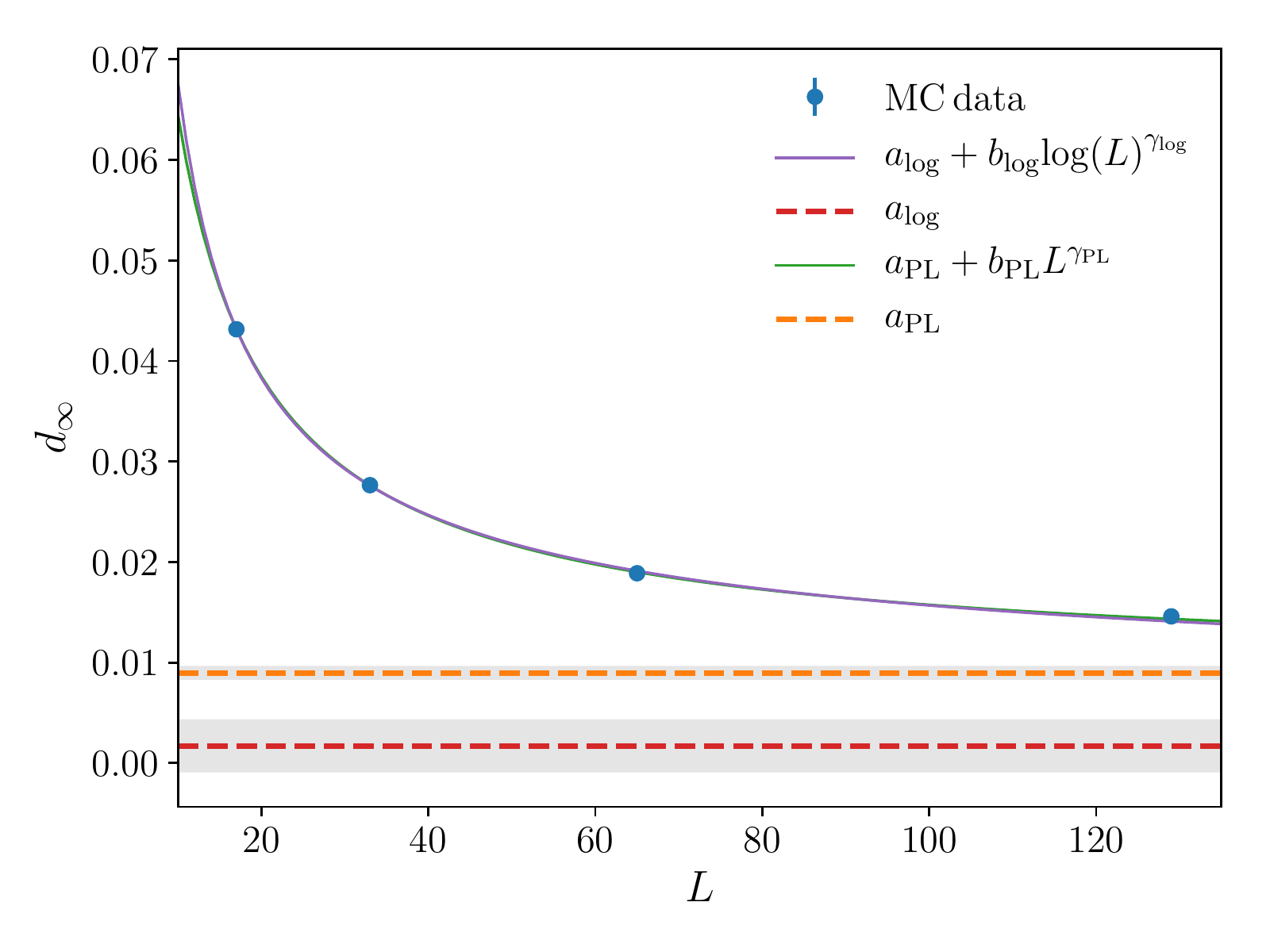}
\caption{Distance $d_\infty$ (see Eq.~\eqref{dinfty}) between the measured $R^{MC}_Q$ and the
CFT prediction as a function of $L$ for $Q=4$. The data can be fitted both with a logarithmic scaling $a_{\log}+b_{\log} \log(L)^{\gamma_{\log}}$
and with a power law model $a_\mathrm{PL}+b_\mathrm{PL}L^{\gamma_\mathrm{PL}}$. We obtain $a_{\log}=(1.7\pm2.6)\cdot 10^{-3}$ (compatible with the theory) and $a_{\mathrm{PL}}=(9.0\pm0.7)\cdot 10^{-3}$;
see main text for a discussion. The shaded area are the errors in the determination 
of $a_{\mathrm{PL}}$ and $a_{\log}$.}
\label{fig_q4}
\end{figure}

\noindent
\textit{$Q=4$ Potts model---} The four-color Potts  model presents different features.
Indeed if we look at Fig.~\ref{fig_diff} 
we see that relative differences (lower part of the plot) are considerably 
larger than on the other cases. The fact that they appear drifting 
toward zero could be reassuring, however, looking at the upper part of the
plot (absolute value of deviation in logarithmic scale), we observe that  convergence, if present, to  Eq.~\eqref{ratio4} is actually very slow. In order to assess
the issue quantitatively, we inspected the behavior of the distance $d_\infty$ in Eq.~\eqref{dinfty}.
Now the choice of the correct model for finite size corrections
becomes crucial. Differently from the previous cases, for the $Q=4$
Potts model \cite{CNS,NS, DGMOPS, BatchPotts} logarithmic  corrections to the scaling are also expected; see also~\cite{Liouville}.
An additional source of complications might be moreover related to the marginality of the boundary operator $\phi_{1,3}$ rightly at $Q=4$. The data, shown in Fig.~\ref{fig_q4}, are reasonably  described ($\chi^2$ per dof approximately $1.5$) by the logarithmic fit 
$a_{\log}+b_{\log}{\log(L)}^{\gamma_{\log}}$ with $a_{\log}=(1.7\pm2.6)\cdot 10^{-3}$, compatible with zero and a slow convergence to the theoretical curve. We also tried to fit the data with a power law decay ($\chi^2$ per dof approximately $0.5$); the two fitting curves in Fig.~\ref{fig_q4}
are barely distinguishable.  The power law fit  yields an  $O(1)$ term ($a_{\mathrm{PL}}=(8.9\pm0.7)\cdot 10^{-3}$) that, although small, is  about one order of magnitude larger than at $Q=1,2,3$ and non-zero within the confidence interval.

We believe that 
both power law and logarithmic
corrections should be present. Disentangling
these two effects is notoriously a difficult task, 
requiring larger system  sizes.
Possible strategies we envisage for settling the question
could be: the search of a geometrical model within the same universality 
class of FK clusters at $Q=4$ but free of logarithmic corrections; a more efficient
sampling of crossing events to access larger sizes, or to
undertake the full analytical study of finite size corrections.
All these paths are certainly worth further investigation.

\section{Conclusions}
In this paper we constructed an universal ratio $R$, see Eq.~\eqref{ratio_r}, that involves four-point boundary connectivities of FK clusters
in the two-dimensional $Q$-color Potts model.
Exploiting lattice duality and conformal symmetry we conjectured an exact expression for $R$ at criticality for any integer values $1\leq Q\leq 4$.
In particular we considerably expanded the study
in~\cite{GV17}, to the three-color and four-color Potts model. 
Remarkably we also provided a conjecture for $R$ in the percolation problem
that corresponds to the limit
$Q\rightarrow 1$. Our theoretical results are plotted  in Fig.~\ref{figrs}.

The percolation case is  particularly interesting  since critical properties are described by a non-unitary CFT with
vanishing central charge. Non-unitary extensions of minimal conformal models~\cite{BPZ} are notoriously hard to  address theoretically and few exact
correlation functions have
been obtained during the years. In particular earlier studies focused on generalizations of Cardy formula~\cite{Cardy_perc}. 

We calculated exactly  four-point functions at $c=0$ of an operator with non-vanishing
scaling dimension and interpreted them as cluster connectivities in critical percolation.
In particular, the sum of connectivities in Fig.~\ref{fig_conn} furnishes a fully explicit example of a logarithmic singularity at $c=0$. 
Consistently with previous analysis~\cite{DJS} and the original proposal~\cite{GL} we also re-obtained the value $b=-5/8$ for the
indecomposability parameter of boundary percolation. This is another direct, although not easily accessible numerically, verification of
the indecomposability parameter $b$
in boundary percolation. We checked extensively our conjectures with high-precision Monte Carlo simulations on a triangular lattice, confirming both universality  of the ratio in 
Eq.~\eqref{ratio_r} and its remarkable agreement with the predictions of conformal invariance for  integers $1\leq Q\leq 3$. For the case $Q=4$ the agreement is confirmed provided the finite size corrections are assumed to
scale logarithmically with the system size~\cite{CNS}; a detailed study of finite size corrections at $Q=4$ is a problem that we left for future work.

\acknowledgments
We are  grateful to John Cardy for
interesting remarks on the draft of this paper, to Mark Van Hoeji for pointing out
the two explicit solutions in Eqs.~(\ref{fl}-\ref{fs}). GG thanks the IIP of Natal for
hospitality. We finally
acknowledge pleasant conversations with Andrea Cappelli, Aldo Delfino,
Giuseppe Mussardo, Raoul Santachiara and Andrea Trombettoni.
\appendix
\section{Frobenius series}
\label{app}
For mathematical details we refer to the classic volume~\cite{Ince}. Given the ODE in Eq.~\eqref{ode} we write its truncated Frobenius series as $G_{\rho}(\eta)=\eta^{\rho}\sum_{k=0}^{N}a_k\eta^{k}$.
We denote by $L$ the action of the differential operator in Eq.~\eqref{ode} then it turns out
\begin{equation}
\label{Frob_series}
L(G_{\rho})=\sum_{k=0}^{N}f_k(\rho)\eta^{\rho+k-1}=0.
\end{equation}
The $k=0$ coefficient in Eq.~\eqref{Frob_series} fixes $\rho$ throughout
\begin{equation}
\label{lin_sist}
f_0(\rho)=a_0\rho(\rho-h)(\rho-3h-1)=0.
\end{equation} 
Once we fixed the normalization $a_0=1$, the remaining  coefficients $a_1, \dots, a_{N}$ in the power series are obtained solving  a linear lower triangular system of equations.
The solution is required in symbolic form to avoid numerical truncation errors when summing the series and can be obtained efficiently up to $N=O(10^5)$ with \texttt{Mathematica}.
In practice, the $N$ equations for the unknowns $a_1,\dots,a_N$  are obtained from \eqref{Frob_series} as 
\begin{equation}
\label{frob_eq}
\frac{1}{l!}\left.\frac{d^l}{d\eta^l}[\eta^{-\rho+1}L(G_{\rho})]\right|_{\eta=0}=f_l(\rho)=0,\quad l=1,\dots,N.
\end{equation} 
The constant vector of the linear system is fixed by the normalization condition $a_0=1$.
However, when there are roots in Eq.~\eqref{lin_sist} that differ by integers, the linear system \eqref{frob_eq} might be inconsistent~\cite{Ince}. In particular let $\rho_1>\rho_0$ two roots of Eq.~\eqref{lin_sist} such that $\rho_1-\rho_0=\nu$ is a positive integer and  denote by $\mathbf{M}^{(\rho_0)}$ the $N\times N$ lower  triangular  matrix associated to Eq.~\eqref{frob_eq} for $\rho=\rho_0$. For simplicity we assume that no other pair of roots are separated by integers. In this case the coefficient $a_{\nu}$ will appear as a free variable (i.e. the element $[\mathbf{M}^{(\rho_0)}]_{\nu\nu}=0$). Furthermore let $\mathbf{M}^{(\rho_0)}_{\nu}$ the matrix obtained replacing the $\nu$-th column of $\mathbf{M}^{(\rho_0)}$ with the constant vector of the linear system.  If the linear system is consistent (i.e. $\text{rk}(\mathbf{M}^{(\rho_0)}_{\nu})=\text{rk}(\mathbf{M}^{(\rho_0)})$) then the particular solution will give the coefficients of the power series of $G_{\rho_0}$, normalized by $a_0=1$.
The kernel of $\mathbf{M}^{(\rho_0)}$ will be spanned by the coefficients of the power series $G_{\rho_1}$ and necessarily $a_0=\dots= a_{\nu-1}=0$; conventionally we choose  $a_{\nu}=1$.

 In this way we generated all the linearly independent power series at $Q=3$ (where $a_3$ is a free variable, $G_0$ is the particular solution and $G_3$ spans the kernel of $\mathbf{M}^{(0)}$) and $Q=4$ (where actually $a_1$ and $a_4$ are both free variables, $G_0$ is the particular solution and the kernel of $\mathbf{M}^{(0)}$ is spanned by $G_1$ and $G_4$). At $Q=1$ and $Q=2$ the linear system associated to Eq.~\eqref{frob_eq} can be inconsistent. For percolation this happens when $\rho_0=0, \rho_1=3h+1=2$  and for Ising when $\rho_0=h=1/2$ and $\rho_1=3h+1=5/2$; in both cases $\nu=2$. When the linear system is inconsistent the particular solution does not exist and it will be replaced by a Frobenius power series with a logarithmic singularity; $G_{\rho_1}$ instead continues to span the kernel of $\mathbf{M}^{(\rho_0)}$ and is free of logarithms.

 The coefficients of the logarithmic solution are determined as follows (again we refer to~\cite{Ince} for a comprehensive discussion that includes the case of repeated roots in Eq.~\eqref{lin_sist}). We introduce a formal power series
\begin{equation}
\label{formal}
G_{\sigma}(\eta)=\eta^{\sigma}\sum_{k=0}^{\infty} b_k(\sigma)\eta^k,
\end{equation}
and solve the linear system in \eqref{frob_eq} choosing $b_0(\sigma)=(\sigma-\rho_0)$. In this way all the coefficients $g_k(\sigma)$  are analytic  in the limit $\sigma\rightarrow \rho_0$. However since~\cite{Ince} $b_k(\rho_0)=0$ for $k<\nu$, it follows that $G_{\rho_0}$ determined through  \eqref{formal} is $b_{\nu}(\rho_0) G_{\rho_1}$ (recall that we chose  $a_{\nu}=1$ for $G_{\rho_1}$ ). To obtain the linear independent solution associated to $\rho_0$ we observe that from the analyticity of the $b_k$'s the differential operator $L$ commutes with the derivative with respect to $\sigma$ and  it turns out
\begin{equation}
L\left(\partial_{\sigma}|_{\sigma=\rho_0}G_{\sigma}\right)=\partial_{\sigma}|_{\sigma=\rho_0}L(G_{\sigma})=\partial_{\sigma}|_{\sigma=\rho_0}(\eta^{\sigma-1} f_0(\sigma)(\sigma-\rho_0))=0.
\end{equation}
The linear independent solution associated to the root $\rho_0$ is then $\tilde{G}_{\rho_0}(\eta)\equiv\partial_{\sigma}|_{\sigma=\rho_0}G_{\sigma}$, i.e.
\begin{align}
\nonumber
\tilde{G}_{\rho_0}(\eta)&=\eta^{\rho_0}\left(\log(\eta)\sum_{k=\nu}^{\infty} b_k(\rho_0)\eta^k+\sum_{k=0}^{\infty}\left.\frac{d b_k}{d\sigma}\right|_{\sigma=\rho_0}\eta^k\right)\\
\label{sollog2}
&\equiv \beta\log(\eta)G_{\rho_1}(\eta)+\sum_{k=0}^{\infty}c_k\eta^{k+\rho_0},
\end{align}
where we defined $\beta\equiv b_{\nu}(\rho_0)$ and $c_k=\left.\frac{d b_k}{d\sigma}\right|_{\sigma=\rho_0}$.
It should be noticed that multiplying $b_0(\sigma)$ by any analytic function $F(\sigma)$ that is $O(1)$ at $\sigma=\rho_0$, the above
procedure produces an equally valid solution of Eq.~\eqref{ode} that corresponds to a linear
combination of Eq.~\eqref{sollog2} and $G_{\rho_1}$; in particular $c_2$ can be arbitrarily redefined. To efficiently generate the power series \eqref{sollog2} we can first determine $G_{\rho_1}$ then fix $c_0=1$ (that in turns fixes $\beta$) and choose $c_2=\left.\frac{d b_2}{d\sigma}\right|_{\sigma=\rho_0}$ with the choice $F(\sigma)=1$. Finally we set up a recursion for the $c_k$ with $k> 2$ requiring Eq.~\eqref{sollog2} to be a solution  of Eq.~\eqref{ode}. 
\section{Recursive formula for the Virasoro conformal blocks}
\label{appzam}
The Virasoro conformal blocks $F_{\rho}^c(\eta)$ can be obtained directly using Al. Zamolodchikov recursive
formula~\cite{Zam_rec}. The formula is conveniently written in terms of the elliptic nome $q=e^{i\tau}$, where the the modulus $\tau$ is related to the
anharmonic ratio by
\begin{equation}
 \tau=i\frac{K(1-\eta)}{K(\eta)},
\end{equation}
where $K$ as in Eq.~\eqref{f} is the complete elliptic integral of the first kind. The Virasoro conformal blocks for an internal field with  dimension
$\rho$ and external legs with dimensions $\{h_i\}$ $(i=1,\dots,4)$ can be explicitly calculated as
\begin{equation}
\label{zamblock}
\mathcal{F}(\eta,c,\rho,\{h_i\})=
 (16q)^{\rho-\frac{(c-1)}{24}}\eta^{\frac{c-1}{24}}(1-\eta)^{\frac{c-1}{24}-h_2-h_3}
 \vartheta_3(q)^{\frac{c-1}{2}-4\sum_i h_i}H(q,c,\rho,\{h_i\}),
\end{equation}
and $\vartheta_3$ is a Jacobi theta function. The function $H$ in \eqref{zamblock}
satisfies th recursion
\begin{equation}
\label{rec_eq}
H(q,c,\rho,\{h_i\})=1+\sum_{r,s}\frac{(16q)^{rs}R_{r,s}(c,\{h_i\})H(q,c,h_{r,s}+rs,
\{h_i\})}{\rho-h_{r,s}(c)}.
\end{equation}
Explicit expressions for $R_{r,s}$ and $h_{r,s}(c)$
(cf. Eq.~\eqref{cc} and Eq.~\eqref{dim}) can be found in~\cite{Zam_rec}
and we will not repeat them here. Since one is interested in generating a series
expansion in $q$ (and ultimately in $\eta$) of Eq.~\eqref{rec_eq} up to order $N$, the level of recursion
is fixed by $rs=N$. Notice that in principle when the internal field
is degenerate, i.e. $\rho=h_{r,s}$ the conformal block might be singular. For all the
non-logarithmic cases examined in this paper, this does not happen
as the corresponding factor $R_{r,s}$ in the numerator of Eq.~\eqref{rec_eq} vanishes as well. The use of the recursive formula for calculating Virasoro conformal blocks for rational values of the central charge $c<1$ has been discussed also in~\cite{Ribault, SantaFoda}.
Since $F^c_{\rho}(\eta)=\mathcal{F}(\eta,c,\rho,\{h,h,h,h\})$,
we used  Eqs.~\eqref{zamblock} and \eqref{rec_eq} to verify (to order $N=10$ in the recursion) all the
identities quoted in Sec.~\ref{sec3}. Special care is needed in the limit $c\rightarrow 1$, to avoid contribution from zero norm states in the identity conformal block (see discussion in the main text).

\section{Solutions at $Q=1$ in terms of hypergeometric functions $_3F_2$}
\label{hyper}
The ODE that is associated to the percolation problem
 is obtained replacing $h=1/3$ in Eq.~\eqref{ode} and reads
\begin{equation}\label{perc_ode}
4(2\eta-1) G(\eta)+ (-6+8 \eta-8 \eta^2) G(\eta)+3 (\eta-1) \eta (2(2\eta-1) G''(\eta)+3 (\eta-1) \eta G'''(\eta))=0.
\end{equation}
\textit{Explicit form for functions entering the ratio $R_{Q=1}$---}
Remarkably, one can formally find two linearly independent hypergeometric solutions~\cite{VH} (one long and the other short)
\begin{align}
\label{fl}
&F_L(\eta)=(\eta (1-\eta))^{4/9} \, _3F_2\left(-\frac{2}{9},-\frac{1}{18},\frac{7}{9};\frac{1}{3},\frac{2}{3};\frac{4}{27}\frac{\left(\eta^2-\eta+1\right)^3}{(1-\eta)^2 \eta^2}\right),\\
\label{fs}
&F_S(\eta)=(1-\eta)^2 \eta^2 \, _3F_2\left(\frac{4}{3},\frac{3}{2},\frac{7}{3};\frac{8}{3},3;4 \eta (1-\eta)\right).
\end{align}
The real and imaginary part of $F_L$, denoted by $F_L^{(R)}$ and  $F_L^{(I)}$ respectively, 
and $F_S$, which is real, constitute an independent basis of the
solutions of \eqref{perc_ode} in the range $0\leq\eta\leq1/2$. 
The computation of $F_L(\eta)$ relies upon the evaluation of a 
$_3F_2$ hypergeometric function with argument in the range $[1,\infty)$
and it is stable for numerical evaluation (in its \texttt{Mathematica}
implementation) so it will be preferred to explicit series expression
of $\tilde{G}_0(\eta)$ and $G_2(\eta)$ (and $G_{1/3}(\eta)$ too) that will be also given.
 This is possible due to the inversion formulas $z\rightarrow 1/z$
 see \cite{3F2inv} that fix the
value assumed by ${}_3F_2$ on the branch cut $(1,\infty)$. All in all the convention results in the function acquiring a value that equals
the one below the branch cut $_3F_2(a_1,a_2,a_3; b_1, b_2, z) \equiv
\lim_{\epsilon\rightarrow 0^+}{}_3F_2(a_1,a_2,a_3; b_1, b_2, z e^{i(2\pi-\epsilon)})$ for real $z>1$.

In order to retrieve the functions $\tilde{G}_0(\eta)$, and $G_2(\eta)$ in Sec.~\ref{sec3}
from the above $F_L(\eta)$, $F_S(\eta)$ within the range $0\leq \eta\leq 1$ we have
to proceed in the following way:
we use linear combinations of $F_L^{(R)}(\eta)$, $F_L^{(I)}(\eta)$, 
and $F_S(\eta)$ for $0\leq\eta\leq1/2$ and of $F_L^{(R)}(1-\eta)$, $F_L^{(I)}(1-\eta)$, 
and $F_S(1-\eta)$ for $1/2\leq\eta\leq1$ and impose the suitable matching condition
at $\eta=1/2$ and normalization.
Define the constants
\begin{equation}
 \alpha_0 = -\frac{3^{7/6} \Gamma\left(-\frac{1}{18}\right) \Gamma\left(\frac{5}{9}\right) 
 \Gamma\left(\frac{7}{9}\right) \Gamma\left(\frac{8}{9}\right)}{2^{13/9} \pi  \Gamma\left(\frac{1}{6}\right)}
\end{equation}
\begin{equation}
 \beta_0=\frac{1}{45} \left(35-12 \log(3)+2 \pi  \left(-\sqrt{3}+\cot\left(\frac{\pi }{9}\right)+
 \cot\left(\frac{2 \pi }{9}\right)+\tan\left(\frac{\pi }{18}\right)\right)\right)
\end{equation}
\begin{equation}
\alpha_2=-\frac{\Gamma\left(-\frac{2}{9}\right) \Gamma\left(-\frac{1}{18}\right) \Gamma\left(\frac{7}{9}\right) \Gamma\left(\frac{8}{3}\right)}
{2^{1/9} \sqrt{3 \pi } \Gamma\left(\frac{1}{3}\right) \Gamma\left(\frac{2}{3}\right) \Gamma\left(\frac{4}{3}\right) \Gamma\left(\frac{7}{3}\right)}.
\end{equation}
The function $\tilde{G}_0(\eta)$ is obtained by imposing the function to be 
continuous, having vanishing first derivatives and continuous second derivative in $\eta=1/2$ and setting the
function to be one for $\eta=0$.
This yields the following expression
\begin{multline}
\label{g0_hyper}
\tilde{G}_0(\eta)=
\alpha_0 \left[\sqrt{3}\sin\left(\frac{2\pi}{9}\right)-\cos\left(\frac{2\pi}{9}\right)\right]F^{(R)}_L(\zeta)+\\
+\alpha_0 \left[-\sin\left(\frac{2\pi}{9}\right)-\sqrt{3}\cos\left(\frac{2\pi}{9}\right)\right]F^{(I)}_L(\zeta)
+\beta_0 F_S(\zeta)
\end{multline}
where $\zeta=\min(\eta,1-\eta)$.\\
In order to reproduce $G_2(\eta)$ we have to impose 
the function to be equal to $F_S(\eta)$ for $0\leq\eta\leq1/2$
and impose continuity of the function and the first two derivatives
at $\eta=1/2$.
The outcome is
\begin{equation}
\label{g1_hyper}
G_2(\eta) = 
  \begin{cases}
    F_S(\eta) & \text{if } 0\leq \eta\leq 1/2,\\
    \alpha_2 F^{(I)}_L(1-\eta)) + F_S(1-\eta) & \text{if } 1/2\leq \eta\leq 1.
  \end{cases}
\end{equation}
The power series expansions of Eqs.~(\ref{g0_hyper}-\ref{g1_hyper}) coincide with Eq.~\eqref{log} and 
Eq. \eqref{2Q1} respectively.
Moreover since we can also calculate explicitly the value of this function
in $\eta=1$ we can fix the normalization constant $A_1$ for the ratio $R_{Q=1}$
given in \eqref{ratioperc}
\begin{equation}
A_1=
\frac{3^{7/6} \pi  \Gamma\left(\frac{5}{9}\right) \Gamma\left(\frac{8}{9}\right) 
\Gamma\left(\frac{7}{3}\right)}{4 \cos(13\pi /18) \Gamma\left(-\frac{2}{9}\right)
\Gamma\left(\frac{1}{6}\right) \Gamma\left(\frac{11}{6}\right)}.
\end{equation}
To have an idea of the power of the derived expression we compare the value
of this expression when $\eta=1/2$ with the series expansion and the numerics. 
The truncated series expansion
(with $10^5$ term) provides the value $R_{Q=1}^{(series)}(1/2)=0.119993\ldots$
while the exact expression gives $R_{Q=1}(1/2)=0.117680185\ldots$.
\newline
\noindent
\textit{Explicit series expression around $\eta=0$---}We provide also 
explicit series expressions derived by working out
the functions $F_L$ and $F_S$
where the argument of the $_3F_2$ functions is 
$\frac{27}{4}\frac{(1-\eta )^2\eta ^2}{\left(1-\eta +\eta ^2\right)^3}$
making them more useful for studying the $\eta\approx 0$ region:
\begin{multline}
\label{g0series}
\tilde{G}_0(\eta)=\left(1-\eta +\eta ^2\right)^{2/3}
g_0\left(\frac{27}{4}\frac{(1-\eta )^2\eta ^2}{\left(1-\eta +\eta ^2\right)^3}\right)+\\
+\frac{1}{45} \left(35-4 i \pi +4 \pi 
\csc\left(\frac{\pi }{9}\right)-12 \log(3)\right)G_2(\eta)
\end{multline}
\begin{equation}
\label{g13series}
G_{1/3}(\eta)=((1-\eta ) \eta )^{1/3} \left(1-\eta +\eta ^2\right)^{1/6}{}_{3}F_2\left(-\frac{1}{18},\frac{5}{18},\frac{11}{18};\frac{1}{6},\frac{7}{6};\frac{27}{4} \frac{(1-\eta )^2 \eta ^2}{\left(1-\eta +\eta ^2\right)^3}\right)
\end{equation}
\begin{align}
\label{g2series}
G_{2}(\eta)&=
((1-\eta ) \eta )^{2} \, _3F_2\left(\frac{4}{3},\frac{3}{2},\frac{7}{3};\frac{8}{3},3;4 \eta (1-\eta)\right)\nonumber\\
&=((1-\eta ) \eta )^{2} \, 
\left(1-\eta +\eta ^2\right)^{-7/3}
{}_{3}F_2\left(\frac{7}{9},\frac{10}{9},\frac{13}{9};\frac{11}{6},2;
\frac{27}{4} \frac{(1-\eta )^2 \eta ^2}{\left(1-\eta +\eta ^2\right)^3}\right)
\end{align}
where we used the additional function
\begin{multline}
g_0(\zeta)=1+\frac{\Gamma \left(\frac{5}{9}\right) \Gamma \left(\frac{8}{9}\right)}
{ \Gamma \left(-\frac{2}{9}\right)\Gamma \left(\frac{1}{6}\right)}
\sum_{k=1}^{\infty}(-\zeta )^k \frac{\Gamma \left(\frac{1}{6}-k\right)
\Gamma \left(-\frac{2}{9}+k\right) }
{(k-1)! k! \Gamma \left(\frac{5}{9}-k\right) \Gamma \left(\frac{8}{9}-k\right)}\nonumber\\
\times\left(\log(\zeta)-i \pi +\psi_0\left(\frac{5}{9}-k\right)+\psi_0\left(\frac{8}{9}-k\right)+\right.\\
\left.+\psi_0\left(-\frac{2}{9}+k\right)-\psi_0(k)-\psi_0(1+k)-\psi_0\left(\frac{1}{6}-k\right)\right)
\end{multline}
and $\psi_0$ is the digamma function. The presence of digamma functions comes not as
a surprise and is analogous to the ones encountered 
when dealing with the logarithmic companion solution to ${}_2F_1(a,b;c;z)$
of the simple hypergeometric differential equation when $c=1,2,3\ldots$
and $a,b\neq 1,2,\ldots, n-1$. A full understanding 
of the relation between the various functions
presented here would entail a better knowledge
of the connection formulas for  $_3F_2$. Unfortunately this theory is not
as developed as the one for the $_2F_1$.

All of these functions indeed reproduce exactly 
the series expansions given in the main text 
in Eq.~\eqref{log}, Eq. \eqref{13Q1}, and Eq. \eqref{2Q1}
but as already noted the series
expression for $\tilde{G}_0$ is not efficient
an the form given in Equation \eqref{g0_hyper} should be preferred.

\bibliographystyle{JHEP}
\providecommand{\href}[2]{#2}\begingroup\raggedright\endgroup
\end{document}